\documentclass[aps,showpacs,twocolumn,superscriptaddress,nofootinbib] {revtex4-1}  

\usepackage{graphicx,color}  
\usepackage{dcolumn}   % needed for some tables
\usepackage{bm}        
\usepackage{amssymb} 
\usepackage{amsmath}
\usepackage{braket}
 \newcommand{\beq}{\begin{equation}}
\newcommand{\eeq}{\end{equation}}
\newcommand{\beqa}{\begin{eqnarray}}
\newcommand{\eeqa}{\end{eqnarray}}
\newcommand{\be}{\begin{equation}}
\newcommand{\ee}{\end{equation}}
\newcommand{\bea}{\begin{eqnarray}}
\newcommand{\eea}{\end{eqnarray}}
\newcommand{\sm}{\mathcal{S}}

\newcommand{\bnabla}{\bar\nabla}

\def\D{\mathcal{D}}

\def\Box{\nabla^2}      
\def\bnabla{\bar \nabla}

\def\bnabla{\bar \nabla}
\usepackage{adjustbox}
\usepackage{standalone}
\usepackage{amsmath}

\usepackage{xcolor}
\usepackage{epsfig}
\usepackage{cancel}
\usepackage{caption}
\usepackage{amssymb}
\usepackage{amsfonts}
\usepackage{rotating}

\usepackage{amsmath}
\usepackage{fancyhdr}

\usepackage{pstricks}
\usepackage{color}
\usepackage{frontespizio}
\usepackage{hyperref}
\hypersetup{
	colorlinks,
	citecolor=green,
	filecolor=black,
	linkcolor=blue,
	urlcolor=black
}
\usepackage{type1ec}
\usepackage[T1]{fontenc}
\usepackage{lettrine}
\usepackage{bbold}
\usepackage{calligra}
\usepackage{tikz}

\usepackage{mathrsfs}
\usepackage{curve2e}
\usepackage{setspace}
\usepackage{indentfirst}
\usepackage{emptypage}
\usepackage[font=small,labelfont=bf,labelsep=quad]{caption} %tabelle e figure
\usepackage{graphicx} %figure
\usepackage{listings} %codici
\usetikzlibrary{patterns}
\usepackage{relsize}
\usetikzlibrary{intersections,positioning}
\usetikzlibrary{decorations.pathmorphing,decorations.markings,arrows,positioning}
\usepackage{braket}
\usepackage{mathrsfs}
\usepackage{stackengine}
\usepackage{calc}
\newlength\shlength
\newcommand\xshlongvec[2][0]{\setlength\shlength{#1pt}%
	\stackengine{-5.6pt}{$#2$}{\smash{$\kern\shlength%
			\stackengine{7.55pt}{$\mathchar"017E$}%
			{\rule{\widthof{$#2$}}{.57pt}\kern.4pt}{O}{r}{F}{F}{L}\kern-\shlength$}}%
	{O}{c}{F}{T}{S}}

\newcommand{\cS}{\mathcal{S}}

\newcommand{\rg}{\sqrt{-g}}

\newcommand{\de}{\delta}

\newcommand{\sdfrac}[2]{\mbox{\small$\displaystyle\frac{#1}{#2}$}}

\IfFileExists{dsfont.sty}
{\usepackage{dsfont}
	\let\mathbb=\mathds
	\newcommand{\id}{\mathds{1}}}
{\typeout{Package dsfont.sty was not found, using alternative macros.}
	\let\mathds=\mathbb
	\newcommand{\id}{\mbox{1 \kern-.59em {\rm l}}}}

\usepackage{slashed}
\usepackage{units}
\usepackage{setspace}
\newcommand{\nn}{\nonumber}
\let\a=\alpha   \let\b=\beta      \let\d=\delta
         
\let\i=\iota        \let\m=\mu
\let\n=\nu

\let\D=\Delta
\let\d=\delta

\newcommand{\pa}{\partial}						% partial derivative sign
\renewcommand{\a}{\alpha}
\newcommand{\bet}{\beta}

\newcommand{\vf}{\varphi}

\def\nbox#1#2{\vcenter{\hrule \hbox{\vrule height#2in
			\kern#1in \vrule} \hrule}}
\def\sq{\,\raise.5pt\hbox{$\nbox{.09}{.09}$}\,}
\def\sqb{\,\raise.5pt\hbox{$\overline{\nbox{.09}{.09}}$}\,}

\newcommand{\na}{\nabla}

\newcommand{\bes}{\begin{subequations}}
	\newcommand{\ees}{\end{subequations}}

\def\nn{\nonumber\\}

\def\Box{\sq}

\numberwithin{equation}{section}

\usepackage{accents}

\makeatletter
\newcommand{\xLine}[2][]{\ext@arrow 0359\Rightarrowfill@{#1}{#2}}
\makeatother
\xdefinecolor{darkgreen}{RGB}{102, 204, 70}
\xdefinecolor{darkblue}{RGB}{0, 0, 153}
\usepackage{amssymb}
\usepackage{pifont}

\begin{document}

\title{\Large \bf  Three-Wave and Four-Wave Interactions in the $4d$ Einstein Gauss-Bonnet (EGB) and Lovelock   Theories\\ }
\vspace{0.1cm}
 \vspace{0.3cm}
\vspace{1cm}
\author{Claudio Corian\`o }
\affiliation{\it Dipartimento di Matematica e Fisica, Universit\`a del Salento \\
and INFN Sezione di Lecce, Via Arnesano 73100 Lecce, Italy\\
National Center for HPC, Big Data and Quantum Computing}
\author{Mario Cret\`i  }
\affiliation{\it Dipartimento di Matematica e Fisica, Universit\`a del Salento \\
and INFN Sezione di Lecce, Via Arnesano 73100 Lecce, Italy\\
National Center for HPC, Big Data and Quantum Computing}
\affiliation{\it Center for Biomolecular Nanotechnologies,\\ Istituto Italiano di Tecnologia, Via Barsanti 14,
73010 Arnesano, Lecce, Italy }
\author{Stefano Lionetti}
\affiliation{\it Dipartimento di Matematica e Fisica, Universit\`a del Salento \\
and INFN Sezione di Lecce, Via Arnesano 73100 Lecce, Italy\\
National Center for HPC, Big Data and Quantum Computing}
\author{Matteo Maria Maglio}
\affiliation{\it  Institute for Theoretical Physics (ITP), University of Heidelberg\\
	Philosophenweg 16, 69120 Heidelberg, Germany}

\begin{abstract}
We derive the conformal constraints satisfied by classical vertices of a (Einstein) Gauss-Bonnet theory around flat space, in general dimensions and at $d=4$ (4d EGB). In $4d$ EGB they are obtained by a singular limit of the integral of the Euler-Poincar\`e density. Our analysis exploits the relation between this theory and the conformal anomaly action, which allows to uncover some interesting features of the GB vertex at cubic and quartic level.  
If we introduce a conformal decomposition of the metric, the resulting theory can be formulated in two different versions, which are regularization dependent, a local one which is quartic in the dilaton field, and a nonlocal one, with a quadratic dilaton. The nonlocal version is derived by a finite redefinition of the GB density with the inclusion of a $(d-4)R^2$ correction, before performing the singular $d\to 4$ limit. In the local version  of the theory, we show how the independent dynamics of the metric and of the dilaton are interwinded by a classical trace identity. Three-gravitational wave interactions can be organised in a nontrivial way by using directly the nonlocal 4d EGB version of the theory. This is possible thanks to the consistency of such formulation - only up to 3-point functions - directly inherited from the conformal anomaly (Riegert) action.  The constraints satisfied by the vertices are  classical, hierarchical Ward identities. At quartic level, similar relations are derived, borrowing from the analysis of the counterterms of the 4T correlators of the conformal anomaly action, as defined by a perturbative expansion. 
For $d\neq 4$ these constraints  hold also for Lovelock actions. They can be extended to higher order topological invariants in such class of theories. 
\end{abstract}

\maketitle

\section{Introduction}
The search for modifications of Einstein's theory of General Relativity (GR) that may 
explain important phenomenological aspects of current cosmology, such as inflation and dark energy, follows several directions. One of them, traditionally, contemplates the inclusion of extra scalar fields into the theory. Such a role is taken by a scalar that drives the metric inflation and couples to all of matter present around the Planck scale, finally decaying into the spectrum of particles, parents of the Standard Model ones. Another modification is the inclusion of a cosmological constant, which fits very well the CMB data within the $\Lambda$CDM model, but underscores a huge hierarchy problem.\\
An interesting class of modified cosmologies are those that include higher powers of the curvature of spacetime, the Riemann tensor, but in a form in which no dimensionful coupling is present in the action and exhibiting equations of motion of the second order, as for the Einstein-Hilbert (EH) action. \\
A nice example of these is the Einstein Gauss-Bonnet (EGB) theory at finite GB coupling, which would be ideal for the study of some of these phenomena, were it not that, in four dimensions, the theory is topological. In string theory in $d=10$ such quadratic corrections get combined in the GB term only in the heterotic case \cite{Duff:1986pq}. It was observed that they are deprived of double poles, generated by the second functional derivative of this term $(\sqrt{g}E^{(2)})^{\mu\nu\rho\sigma}$, from the quadratic metric fluctuations around flat space $(\sim O(h^2))$ \cite{Zwiebach:1985uq}. \\
We recall that topological contributions in the form of either the Einstein-Hilbert (EH) action at $d=2$ 
\beq
V_{EH}(g,d)\equiv\mu^\epsilon\int d^d x \sqrt{g} R,
\eeq
$\epsilon=d-2$,
or the Gauss Bonnet action (GB) at $d=4$, define evanescent terms in the equations of motion of gravity. In $d=2$ the EH action itself is metric independent.\\
 Evanescent terms can be turned into dynamical contributions by performing a singular limit on the corresponding coupling constant, which are dimensionless. This features is commonly present and held into account in the context of conformal anomaly actions, but recently it has been reproposed in a purely classical context \cite{Glavan:2019inb}.
 In the case of $d=2$, the theory is rendered dynamical by replacing the EH action by the regulated action 
 \cite{Mann:1992ar}
 \beq
 \label{eh2}
 \sm_2=\lim_{d\to 2}\frac{\int d^d x \left(V_{EH}(g,d)- V_{EH}(\bar g,d)\right)}{d-2},
 \eeq
 where one introduces a conformal decomposition of the metric 
 \beq
 g_{\mu\nu}=\bar{g}_{\mu\nu}e^{2 \phi}
 \eeq
 in terms of a dilaton (Weyl) factor $\phi$ and a fiducial metric $\bar{g}$, with $V_{EH}(\bar g,d)$ being a subtraction that allows a finite limit. The limiting procedure allows to generate special forms of dilaton gravities, which are closely related to conformal anomaly actions, and are of Horndeski type. The differences among the possible realizations of such actions are related to the ways in which the subtractions are included in the regularization of such $V_{EH}(g,d)/\epsilon\to 0/0$ contributions as $\epsilon\to 0$.  \\
 The method, well-known both in the case of 2-D gravity and of conformal anomaly actions \cite{Matsumoto:2022fln,Coriano:2013nja,Ferreira:2017wqz,Elvang:2012st}, where the same procedure is applied to the counterterms in the dimensional regularization (DR) of the theory, has recently regained significant attention, for offering, possibly, a way to evade Lovelock' s theorem 
\cite{Lovelock:1971yv} in a purely classical framework \cite{Glavan:2019inb}. \\
Eq. \eqref{eh2} identifies the ordinary Wess-Zumino (WZ) form of the action. Different subtractions, defined either in $d$ dimensions - as in \eqref{eh2} - or at $d=4$, allow to include or exclude extra - Weyl invariant - terms. 
It has been pointed out that this arbitrariness is a possible way to account for the difference between topological and non topological anomalies, or anomalies of types A and B \cite{Coriano:2023sab}. Both types of anomalies appear in the trace of correlation functions involving stress energy tensors $(T)$, but the first are not 
associated with the breaking of scale invariance.\\
Similarly to the $d=2$ case, in the case of the 4d GB theory, the coupling $g_s(d)$ becomes singular as $d\to 4$, while the integral of the Euler-Poincar\`e density becomes topological, generating again a $0/0$ contributions that requires a subtraction. The result is a finite, non topological  action, whose structure depends on the subtraction 
\cite{Coriano:2022ftl}. \\  
 The $d\to 4$ limit is purely geometrical, but involves additional scales, coming from the dependence of the fields on the extra $(d-4)$ coordinates of the manifold, and borrows its features from dimensional regularization (DR) in flat Minkowski space. As a theory, it is unrelated to the regularization of some quantum corrections, as is the case of the conformal anomaly action, derived by integrating out a conformal matter sector. However, it is a well-defined variant that opens the way to new interesting developments. It is clear that such behaviour can be classified as a strongly coupled one, geometrically induced and with a dimensional reduction of the field dynamics. It can be envisioned for every topological term of a general Lovelock theory.\\ 
Lovelock's theorem states that, at $d=4$, the only gravitational action that generates second order equations of motion is the EH action, plus a cosmological constant
\beq
\sm_{EH}=\int d^d x \sqrt{g}( M_P ^2 R +2 \Lambda).
\eeq
Its generalization to higher dimensions takes the form \cite{Lovelock:1971yv}
\begin{align}   
&{\cal L}^{(n)}={n!\over 2^{n/2}} \delta^{[\mu_1}_{\nu_1} \cdots \delta^{\mu_n]}_{\nu_n}R_{\mu_1\mu_2}^{\ \ \   \ \ \nu_1\nu_2}R_{\mu_3\mu_4}^{\ \ \ \ \ \nu_3\nu_4}\cdots\nonumber \\
&\cdots R_{\mu_{n-1}\mu_n}^{\ \ \ \ \ \ \ \nu_{n-1}\nu_n},\ \ \ n=0,2,4,\ldots \,.
 \label{lovelock1} 
 \end{align}
$n=0$ identifies the cosmological constant, $n=2$ the EH action and $n=4$ the GB density. Given the antisymmetrization present in its definition, this is nonzero only in specific dimensions. Once the indices are contracted, the resulting density and its coordinate dependence can be extended to any dimension. 
The reduction of the action to the topological dimension $d=4$ for E ($E_4$) is investigated by an embedding of the metric into the extra $(d-4)$ dimensions, a procedure which is not unique \cite{Coriano:2022ftl}. 
The case $n=4$, with $d=4$, defines the GB density 
\begin{align}
\label{ffr}
V_E(g,d)\equiv &\mu^{\varepsilon} \int\,d^dx\,\sqrt{-g}\,E , 
\end{align}
where $\mu$ is a renormalization scale  and $E$ is the integrated Euler-Poincar\`e density 
\beqa
\label{GB1}
 E& =& R^2 - 4 R^{\mu \nu} R_{\mu \nu} + R^{\mu \nu \rho \sigma} R_{\mu \nu \rho \sigma},
\eeqa
whose inclusion modifies the EH action just by boundary contributions, since in an ordinary EGB theory
\beq
\mathcal{S}_{EGB}=S_{EH} + g_s V_E,
\label{first}
\eeq
the GB term is evanescent at $d=4$. Its contribution to the gravitational equation of motion  
 \beq
\frac{1}{\kappa}\left(R_{\mu\nu}-\frac{1}{2}g_{\mu\nu}R+\Lambda_{0}g_{\mu\nu}\right)+ g_s(V_{E}(d))_{\mu\nu}=0,
\label{GB2}
\eeq
 explicitly given by
\begin{align}
\label{ep3}
&V_E^{\mu\nu}\equiv \frac{\delta V_E}{\delta g_{\mu\nu}}= \rg \biggl ( \frac{1}{2} g^{\mu \nu} E_4 - 2 R^{\mu \alpha \beta \gamma}R^\nu_{\alpha \beta \gamma} +\nonumber \\
& 4 R^{\mu \alpha} R^\nu_{\ \alpha} + 4 R^{\mu \alpha \nu \beta} R_{\alpha \beta} - 2 R R^{\mu \nu} \biggl ),
 \end{align}
vanishes at $d=4$ if we use \eqref{GB1}. In $d>4$,  $V_E$ is not a boundary term, and is indeed contemplated by Lovelock' s theorem as a possible modification of the EH action (see 
\cite{Edelstein:2014dje, Charmousis:2014mia}). 
We are going to investigate the conformal constraints associated 
with this term in $d$ dimensions, that have not been investigated before. A similar singular limit can be performed at $d=6$ for the topological invariant $E_6$, cubic in the curvature, extending the strategy discussed in this work.\\ As we have already discussed in the introduction, such constraints are a natural consequence of the role played by such terms in the context of conformal anomaly actions in every even dimensions. \\
As mentioned, when expanded around a flat spacetime 
$g_{\mu\nu}=\eta_{\mu\nu} +\kappa h_{\mu\nu}$, the operators of highest derivatives $(\Box^2)$ of the GB action, contributing to the quadratic term in the action ($h\Box^2 h$), cancel out, showing that the theory is free of ghosts. 
We recall that at quadratic order, the contribution to an action containing the Riemann tensor and its contractions, with arbitrary combinations, is affected by a propagator with double poles in the form 
\begin{align}
\label{exE}
&\int d^d \sqrt{g} \left((R_{\mu\nu\rho\sigma})^2 + a_1 (R_{\mu\nu})^2 + a_2 R^2\right)=\nonumber \\
&\frac{1}{4}\int d^d x \sqrt{g}\left( 
(a_1 + 4) h_{\mu\nu} \Box^2 h_{\mu\nu} +(a_2-1)h\Box^2 h \right) +\nonumber \\
& O(h^3)
\end{align}
that vanish if $a_1$ and $a_2$ are chosen to reproduce the Euler-Poincar\`e density. \\
For general metric background, the analysis of the behaviour of such actions can be performed starting from a conformal decomposition 
\beq
\label{cd}
g_{\mu\nu}=\bar{g}_{\mu\nu}e^{2 \phi}
\eeq
and eliminating the dilaton using its expression in terms of the entire metric $g$. 
The theory becomes nonlocal if the dilaton is removed from the spectrum, as suggested for the conformal anomaly action in \cite{Mazur:2001aa}. One relies on integrable conformal decompositions, such as the one discussed in 
\cite{Barvinsky:1995it}, recently investigated in connection with the perturbative hierarchy of the conformal Ward identities for a specific 4-point function in \cite{Coriano:2022jkn}. Different nonlocal forms of such actions are possible, which differ - rather nontrivially- by different possible inclusions of Weyl invariant terms.  
 As in the $d=2$ case, in order to bring the dynamics of $V_E$ down to $d=4$ from $d>4$, bypassing its evanescence, one can perform a similar singular rescaling of the coupling 
\beq
\label{re}
g_sV_E\to g_s(d) V_E\qquad  g_s(d)\equiv \frac{g_s\mu^\epsilon}{d-4} 
\eeq
 in order to remodulate $g_s V_E$ as a $0/0$ contribution. 
Obviously, the definition of the 
$4d$, $(d=4)$ singular limit of the GB theory, requires a specific compactification, which depends on the underlying geometry and is, in general, affected by extra Kaluza-Klein modes.  
We are essentially performing an infinite coupling limit ($g_s\to g_s(d)$) on the GB term, as we approach the dimension at which the GB contribution is topological.  
These variant theories should be seen as classical modifications of the EH action that resolve the evanescence of a certain topological term in specific even dimensions.\\
Compared to the ordinary Lovelock's classification of pure gravity actions yielding equations of motion of second order, this procedure is essentially new.  As mentioned, it can be performed in any even dimension, starting with $d = 2$.\\
In this work we are going to provide the expression of the nonlocal EGB action expanded up to quartic order (4-graviton vertex) in the fluctuations around a classical metric background. 
The result is derived elaborating on various previous analysis of the conformal anomaly actions, adapted and simplified for $4d$ EGB theories. Notice that the conformal Ward identities (CWIs) derived for the 
$\sm_{WZ}$ theory carry the same structure of the anomalous CWIs characterizing the quantum anomaly action \cite{Coriano:2021nvn}, where vertices of the $V_E$ term defined in \eqref{ffr}, obtained by differentiating this functional, are constrained by the fundamental symmetries. \\
The main difference between a classical and a quantum approach lays in the fact that the constraints obtained by the procedure either apply to classical vertices - for  $4d$ EGB theories - or to quantum averages of correlation functions, if the analysis is performed in a quantum context. In this second case, the classical action in the functional integral    
$S_o(\chi_i,g)$, where the $\chi_i$'s are generic conformal field, is decomposed in terms of the two components $\bar{g}$ and $\phi$ of \eqref{cd}, corresponding to the fiducial metric and the dilaton field.  A similar decomposition can be introduced for the GB term, with equations of motion that are constrained by the "anomaly" of $V_E$, defined below in \eqref{epx}, which is Weyl variant for general $d$. 

\subsection{Singular rescalings and finite subtractions}
The singular rescaling of the coupling can be applied to any topological term, such as $E_4, E_6$ and so on. In practice, the method is sufficient in order to regulate the $0/0$ limit of the $d=4$ action, though the result, as we have mentioned, depends on the geometry of the compactification. In practical terms, the resulting actions are usually simplified, by  neglecting the dependence of the metric on the extra coordinates while they reduced them to $d=4$. This is a procedure that, even if not stated 
explicitly in the literature, is essentially based on dimensional reduction (DRed), in the form described in \cite{Coriano:2022ftl}. Conformal anomaly effective actions, to which $4d$ GB models are related, are derived by a similar procedure, applied to the GB ($V_E$) and the Weyl tensor squared ($V_{C^2}$) counterterms. The latter, in the case of a $4d$ EGB,  is unnecessary.  \\     
The topological evanescence of the $V_E$ contribution is lifted by the procedure, but some ambiguities are encountered, due to
the non-unique choice of the background metric against which $ V_E $ is calculated, which remains an indetermination of the method \cite{Coriano:2022ftl}.  \\
The derivation of the geometric effective action depends on the specific choice of the fiducial metric and of the subtraction term, here identified in the form of a Wess-Zumino action \cite{Coriano:2022knl} via the conformal decomposition, although other subtractions are possible.  
A discussion of this point can be found in \cite{Coriano:2022ftl}. \\
The regularization of the action that results from \eqref{re} is not uniquely defined, since the DRed procedure is naturally affected by an integration cutoff. 
The singular limit of the GB term is investigated by a Weyl rescaling of this term in $d\neq 4$, which introduces 
a dilaton in the spectrum, and the $\epsilon=d-4\to 0$ expansion is performed afterwards, accompanied by the DRed procedure.

\subsection{Content of this work}
The goal of our work is to identify the constraints satisfied by the classical vertices of the theory, obtained by removing the dilaton from the spectrum, and resorting to a nonlocal description of the 4d GB theory. The local version of such theory, which is given by dilaton gravity, therefore, is replaced by a nonlocal theory  when a finite (classical) renormalization of the GB interaction, proportional to an $R^2$ term,  
is added to the usual GB action,  which is allowed by the singular limit .  \\
We identify the constraints satisfied by the vertices of such nonlocal action once its expression is expanded around flat space. 
These correspond to a set of Ward identities which are naturally satisfied by the Weyl-variant part of a renormalized anomaly action, due to the similarity between such action, which describes the Weyl/conformal anomaly, and the $4d$ EGB theory. We provide a description of such vertices up to quartic order in the fluctuations around a flat background. The conformal constraints satisfied by the vertex derived from $V_E(g,d)$ are valid in $d$ dimension due to the fact that Weyl variation of this term given by \eqref{epx} is exactly linear in $(d-4)$. \\
At $d=4$ the evanescence of the term, as already pointed out, is removed by the inclusion of a subtraction, corresponding to a classical renormalization, and the conformal constraints remain valid once we replace $V_E(g,d)$ by the regulated vertex $\hat{V}'_E$, which describes the WZ form of the action.  \\
Section 4 contains a first principle discussion of the constraints on the equations of motion found once we perform a conformal decomposition, showing that they are a rigorous consequence of symmetry \eqref{sym} .
The latter is broken by the subtraction term introduced in the definition of $\hat{V}'_E$. Also in this case, the corresponding constraint, given in \eqref{fin}, is naturally borrowed from the case of the conformal anomaly actions \cite{Coriano:2022ftl}.

\section{The local EGB theory and the nonlocal action}
The correctly regulated theory takes the form of a Wess-Zumino (WZ) action, which depends on the regularization procedure and the treatment of the dilatonic field $(\phi)$. The $0/0$ regularization follows closely the 2D case, where the Einstein-Hilbert term is also topological, and the limit is performed by redefining the coupling as $\alpha\to \alpha/(d-2)$.
  
In general, dilaton effective actions may contain  solutions with the conformal factor that needs to be stabilized around a certain scale $f$. Such a scale is the conformal breaking scale. The scale ($f$) is required in order to redefine the dimensionless conformal factor of the metric $g$ in the conformal decomposition \eqref{sym}. The local shift symmetry, which allows to identify a fiducial metric and the dilaton field, via the transformation 
\beq
\label{sym}
\phi\to \phi -\sigma , \,\,\bar{g}_{\mu\nu}\to \bar{g}_{\mu\nu} e^{2\sigma}
\eeq
with $\sigma=\sigma(x)$, is indeed broken by the regularization of the Lagrangian in the $0/0$ limit. 
This issue is not present in the nonlocal action, since $\phi$ can be eliminated in terms of the entire metric, but, as we have already mentioned, one needs to perform an additional finite renormalization of the action \cite{Mazur:2001aa} in order to reduce the equations of motion for $\phi$ to a linear form. 

A EGB theory is not uniquely defined in such a singular limit, due to several issues, related to the selection of the background metric and to the regularization procedure that it is invoked. The $V_E$ (GB) term can be expanded around $d=4$ in several ways. One possibility is defined by the ordinary DR-like procedure 

\begin{equation} 
\label{expand1}
\frac{1}{\varepsilon}V_{E}(g,d)=\frac{\mu^{\epsilon}}{\epsilon}\left( V_{E}(g,4) + \epsilon 
V_{E}'(g,4) +O(\varepsilon^2) \right),
\end{equation}
in terms of a single metric $g$, 
implicitly defining the GB part of the EGB action in the form 
\beq
\label{oone1}
V'_E=\frac{1}{\epsilon}\left( V_E(g,d)-V_E(g,4)\right).
\eeq
Note that the subtraction term $V_E(g,4)$ obviously does not contribute to the dynamics, for being topological, and amounts just to a constant being added to the action, since
\beq
V_E(4)=\int d^4 x \sqrt{g} E=4\pi \chi_0(\mathcal{M}),
\eeq
where $\chi_0(M)$ is the Euler-Poincar\`e characteristic of a manifold $M$.
Therefore, the evanescence of the GB contribution $V_E$ is related to the fact that at $d=4$ its variation is zero, together with all the classical vertices derived from its functional differentiation 
\beq
\label{es}
V_E^{\mu_1\nu_1\ldots \mu_n\nu_n}=\frac{\delta^n V_E(4)}{\delta g_{\mu_1\nu_1}\ldots  \delta g_{\mu_n\nu_n}}.
\eeq 
The finiteness of the contributions generated by the renormalized vertices $1/\epsilon V_E^{\mu_1\nu_1\ldots \mu_n\nu_n}(d)$ is therefore related to the $O(\epsilon)$ behaviour of \eqref{es} as $\epsilon\to 0$, as we will discuss next. This generates a finite EGB  theory of the form 

\beq
\mathcal{S}_{EGB}=S_{EH} +V'_E,
\label{first1}
\eeq
where $V'_E$ is bound to satisfy the constraint 
\beq
\frac{\delta}{\delta\phi} V'_E=\sqrt{g} E,
\label{he}
\eeq
as recognized in the conformal anomaly effective action. 
Alternatively, the finite action could be defined in the Wess-Zumino (WZ) form 

\beq
\label{WZ}
\mathcal{S}_{WZ}\equiv \hat{V}'_E= \lim_{d\to 4}\frac{1}{\epsilon} \left(V_{E}(\bar{g} e^{2 \phi},d) -V_E(\bar{g},d)\right)
\eeq
that differs from \eqref{oone1} by Weyl invariant terms
\beq
\mathcal{S}_{EGB}=S_{E.H} +\sm_{WZ}.
\label{first2}
\eeq
The different EGB actions that can be generated in the $d\to 4$ limit are all associated with the treatement of the $V_E$ term, a procedure that should be completely defined in DR and with the choice of a specific fiducial metric $\bar{g}$. This would correspond to the choice of a specific scheme, as usually done in Minkowski space. 
Note that contracting \eqref{ep3} with $2 g^{\mu\nu}$ gives the relation
\beq
\label{epx}
2 g_{\mu\nu}\frac{\delta}{\delta g_{\mu\nu}}\int d^d y \sqrt{-g} E(y)=\epsilon \sqrt{g}E(x),
\end{equation} 
which is at the core of \eqref{he}, since the subtraction term $V_E(\bar g, d)$ is Weyl independent.  Such subtraction is essential in order to generate a $0/0$ limit of the topological term and obtain, henceforth, a finite action. This Weyl variation is an exact property of the $V_E$ terms, therefore valid to all orders in $\epsilon$. We will come back to it in a next section, when discussing its implication in the context of Lovelock theory. 

\eqref{epx} is identified using the scaling relation  
\begin{align}\label{GaussB}
&\rg E=\sqrt{\bar g} e^{(d-4)\phi}\biggl \{ \bar E+(d-3)\bar\nabla_\mu \bar J^\mu(\bar{g},\phi)+ \nonumber \\ 
&(d-3)(d-4)\bar  K(\bar{g},\phi)  \biggl \},
\end{align}
where we have defined
\begin{align} \label{GBexJ}
&\bar J^\mu(\bar{g},\phi)=8\bar R^{\mu\nu}\bar\nabla_\nu\phi-4\bar R\bar \nabla^\mu\phi +\nonumber \\
&4(d-2)(\bar\nabla^\mu\phi\bar \Box \phi-\bar \nabla^\mu\bar\nabla^\nu\phi\bar \nabla_\nu\phi+\bar\nabla^\mu\phi\bar\nabla_\lambda\phi\bar\nabla^\lambda\phi),
\end{align}
\begin{align} \label{GBexK}
&\bar K(\bar{g},\phi)=4\bar R^{\mu\nu}\bar\nabla_\mu\phi\bar\nabla_\nu\phi-2\bar R\bar\nabla_\lambda\phi\bar\nabla^\lambda\phi+\nonumber \\
&4(d-2)\bar\Box\phi\bar\nabla_\lambda\phi\bar\nabla^\lambda\phi+(d-1)(d-2)(\bnabla_\lambda \phi\bnabla^\lambda \phi)^2,
\end{align}
that allows to perform the expansion in $\epsilon$ of the form 

\begin{align}
\label{red1}
&V_E(g,d)=\int d^d x \sqrt{\bar{g}}\left(\bar E + \bar\nabla_M \bar J^M \right) +\nonumber \\
&\epsilon\int d^d x \sqrt{\bar g} \phi\left( \bar E + \bar\nabla_M \bar J^M \right) +\epsilon \int d^d x \sqrt{\bar g} K.
\end{align}
The scheme dependence of the regularization comes as a last step, when the integrals present in \eqref{WW} are reduced to 
$d=4$ from general $d$ dimensions. This can can be obtained by introducing a cutoff ($L$) in the extra dimensions in the form 

\bea \label{WG in generic case}
&\frac{1}{d-4}V_E(g,d) =\frac{1}{\epsilon} \left({L}{\mu}\right)^{\epsilon}\int d^4x \rg \  {}_4 \bar{E}+\nonumber\\
&+ \left({L}{\mu}\right)^{\epsilon}\int d^4x \rg\ \Big[\phi {}_4\bar{ E}-(4 {} G^{\mu\nu}(\bar\nabla_\mu\phi\bar\nabla_\nu\phi)  +\nonumber \\
&2(\nabla_\lambda \phi \nabla^\lambda \phi )^2 +4\Box\phi \nabla_\lambda \phi \nabla^\lambda \phi ) \Big],
\eea
where all the terms in the integrands are 4-dimensional and $L$ is a space cutoff in the $d-4$ extra dimensions. $L^{\epsilon}$ is the volume of the extra space. Taking the $\epsilon\to 0$ limit and the conformal separation $g_{\mu\nu}=\bar g_{\mu\nu}e^{2 \phi}$ for the fiducial metric, we finally derive the expressions
\begin{align}
\label{rdef}
&\hat{V}'_E( g, \phi)\equiv\sm_{WZ}=\frac{1}{\epsilon}\left(V_E(g,d)-V_E(\bar g,d)\right)=\nonumber\\
&\int d^4x \rg \Big[\phi {}_4 E-(4 {} G^{\mu\nu}(\bar\nabla_\mu\phi\bar\nabla_\nu\phi)+2(\nabla_\lambda \phi \nabla^\lambda \phi )^2 +\nonumber \\
&4\Box\phi \nabla_\lambda \phi \nabla^\lambda \phi ) \Big].
\end{align}
It is easy to show that the use of the regularization in the form given above, by subtracting $V_E(\bar g,d)$ in 
$d$ dimensions - rather than at $d=4$ - is eliminating some Weyl invariant terms \cite{Coriano:2022ftl}.
The local action given above is quartic in $\phi$, and its structure depends on the chosen fiducial metric.\\
In summary, it is possible to define a consistent procedure for the extraction of the effective action at $d=4$, from the singular limit of a topological term. The approach can be performed in $d$ dimensions by a 1) rescaling of the topological density using 
\eqref{GaussB}, with a metric which is d-dimensional.  This implies, obviously, that the dilaton field carries dependence on the extra dimensions. At the last stage,  2) we dimensionally reduce the fields, by allowing only the zero mode of $\phi$  to survive the compactification procedure, while the extra components of the metric are assumed to be flat. The cutoff $L$ in the size of the extra dimensions is introduced in order to guarantee the convergence of the integral $V_E$ in the $d\to 4$ limit. Finally, 3) we subtract the same term expressed only in terns of the fiducial metric, performing the limit.\\
 As shown above, the subtraction can be performed either as in \eqref{oone1} or as in \eqref{WZ}, the difference between the two being given by Weyl invariant terms, described in \cite{Coriano:2022ftl}.
The $\sm_{WZ}$ action, which identifies a contribution that we have also called $\hat{V}'_E$ in \eqref{WZ}, will define our starting action.  
\subsection{The $4d$ EGB + $R^2$ theory}
We briefly summarize the steps that take to the nonlocal action.
One may proceed  by introducing a finite renormalization/extension 
of the topological term, in order to derive a different version of $\sm_{WZ}$, which is quadratic in $\phi$, rather than quartic. This is obtained by extending the topological term at $O(\epsilon)$ in the form \cite{Mazur:2001aa}

\beq
E_{ext}=E_4 +\epsilon\frac{R^2}{2 (d-1)^2},
\label{eee}
\eeq
and the singular limit performed on the functional
\beq
\label{ss1}
\tilde{V}_E=\int d^d x \sqrt{g}E_{ext}. 
 \eeq
The effective action is then defined similarly to \eqref{fin}, with $\sm_{WZ}$ in \eqref{rdef} now redefined by the inclusion of  \eqref{eee}
\beq
\label{MWZ}
\tilde{\mathcal{S}}_{WZ}= \frac{1}{\epsilon} \left(\tilde{V}_{E}(\bar{g} e^{2 \phi},d) -\tilde{V}_E(\bar{g},d)\right)
\eeq
induced by this additional finite modification of the action.
A direct computation, using the rescaling formula for $R^2$ 
\begin{align}
&\sqrt{g}R^2=\sqrt{\bar g}e^{\epsilon \phi}\biggl( \bar R -2 (d-1) \bar\Box \phi -\nonumber \\
&(d-1)(d-2) \bar\nabla_\lambda\phi\bar\nabla^\lambda\phi\biggl)^2\biggl )
\end{align}
gives, after an expansion at $O(\epsilon)$
\begin{align}
&\frac{1}{\epsilon}\int d^d x \sqrt{g}E_{ext}=\frac{1}{\epsilon}\int d^4 x\sqrt{\bar g} \biggl(  \bar E +  \bar\nabla\cdot \bar J \biggl) +\nonumber \\
&\int d^4 x\sqrt{\bar g}\phi\biggl(\bar E +  \bar\nabla\cdot \bar J \biggl) \nonumber \\
&+\int d^4 x\sqrt{\bar g}\biggl( \bar K  + \frac{1}{2(d-1)^2}\biggl[ \bar R -2 (d-1) \bar\Box \phi -\nonumber\\
&(d-1)(d-2) \bar\nabla_\lambda\phi\bar\nabla^\lambda\phi\biggl]^2 \biggl).\nonumber \\
\end{align}
The expression can be simplified by some integration by parts and the omission of boundary terms. Explicitly, one uses
\begin{align}
&\phi \bar\nabla\cdot \bar J=-8 \bar R^{\mu\nu}\bar\nabla_\mu\phi \bar\nabla_\nu\phi
+4\bar R\bar\nabla^\mu\phi\bar\nabla_\mu\phi-\nonumber \\
&4(d-2)\bar\nabla_\mu\phi \bar\nabla^\mu\phi
\bar\Box\phi +4(d-2) \bar\nabla^\mu\bar\nabla^\nu\phi \bar\nabla_\nu\phi\bar\nabla_\mu\phi\nonumber\\
&-4 (d-2)(\bar\nabla_\mu\bar\nabla^\mu\phi)^2 +\textrm{b.t.}
\end{align} 
and 
\beq
 \bar\nabla^\mu\bar\nabla^\nu\phi \bar\nabla_\nu\phi\bar\nabla_\mu\phi=-\frac{1}{2}
\bar\nabla^\mu\phi\bar\nabla_\mu\phi\bar\Box\phi +\textrm{b.t.}
\eeq

(where $\textrm{b.t.}$ indicates the boundary terms).
 This gives the modified relation

\beq
\delta_\phi\int d^d x \sqrt{g} E_{ext}=\epsilon \sqrt{g}(E_{ext}- \frac{2}{d-1} \Box R) 
\eeq
which can be used in \eqref{ss1} to give

\beq
\frac{\delta}{\delta\phi}\frac{1}{\epsilon}\tilde{V}_E(g_{\mu\nu},d)= \sqrt{g}\left(E-\frac{2}{3}\Box R +
\epsilon\frac{R^2}{2(d-1)^2}\right).
\eeq
Therefore from \eqref{MWZ} we derive the Weyl variation
 \beqa
 \frac{\delta \mathcal{S}^{WZ}}{\delta\phi}&=&\alpha\sqrt{g}\left(E-\frac{2}{3}\Box R \right)\nonumber \\
&=&\alpha\sqrt{\bar g}\left(\bar E-\frac{2}{3}\bar \Box\bar R + 4 \bar\Delta_4 \phi\right).
\label{solve}
\eeqa

Note that the redefinition of the GB density $(E\to E_{ext})$ allows to re-obtain a rescaling of the combination $E -2/3 \Box R$
as in $d=4$, where $\Delta_4$ is the fourth order self-adjoint 
operator. This is conformal invariant 
\beq
\Delta_4 = \na^2 + 2\,R^{\mu\nu}\na_\mu\na_\nu - \frac23\,R{\Box}
+ \frac13\,(\na^\mu R)\na_\mu\,
\label{120}
\eeq
and satisfies the relation
\beq
\sqrt{-g}\,\D_4\chi=\sqrt{-\bar g}\,\bar{\D}_4 \chi,\label{point2}
\eeq
if $\chi$ is invariant (i.e. has scaling equal to zero) under a Weyl transformation. This brings the action $\tilde{\mathcal{S}}_{WZ}$ to the form
\begin{equation}
\tilde{\mathcal{S}}_{WZ}= \alpha\int\,d^4x\,\sqrt{-\bar g}\,\left\{\left(\overline E - {2\over 3}
\bar{\Box} \overline R\right)\phi + 2\,\phi\bar\Delta_4\phi\right\}.\,
\label{WZ2}
\end{equation}
The elimination of $\phi$ can be performed quite directly. Using the currents 
\begin{align}
 &J(x)=\bar{J}(x) + 4 \sqrt{g}\Delta_4\phi(x),\qquad     \bar J(x)\equiv \sqrt{\bar g}\left( \bar E-\frac{2}{3}\bar \Box \bar R\right), \nonumber\\
 &  J(x)\equiv \sqrt{ g}\left(  E-\frac{2}{3} \Box  R\right) 
 \label{inv}
 \end{align}
 and the quartic Green function of $\Delta_4$
\begin{equation}
(\sqrt{-g}\,\D_4)_xD_4(x,y)=\d^4(x,y).\label{point4}
\end{equation}
We can invert \eqref{inv}, obtaining 
\begin{equation}
\label{onshell}
\phi(x)=\frac{1}{4}\int d^4y\,D_4(x,y)(J(y)- \bar{J}(y)).
\end{equation}
The expression of $\phi$ in terms of the entire metric $g$ is what defines a conformal decomposition of the metric, which in this case is integrable, in the sense that we can express $\phi$ covariantly.  
 $\tilde{\sm}_{WZ}$ can be obtained by solving the equation 
\beq
\frac{\delta \tilde{\mathcal{S}}_{WZ}}{\delta \phi}=J,
\eeq
clearly identified in the form
\beq
\tilde{\sm}_{WZ}=\int d^4 x \left(\bar J \phi + 2\phi(\sqrt{g} \Delta_4) \phi\right).
\eeq
At this stage it is just matter of inserting the expression of $\phi$ given by \eqref{onshell} into this equation to obtain 
the WZ action in the form
\beq
\tilde{\sm}_{WZ}=\sm_{anom}(g)- \sm_{anom}(\bar g),
\label{WW}
\eeq
with 
\beq
\sm_{anom}(g)=\frac{1}{8}\int d^4 x d^4 y J(x) D_4(x,y) J(y),
\eeq
and a similar expression for $\sm_{anom}(\bar g)$. 
Using the explicit expression of $\phi$, and including the contribution from the rescaled $C^2$ term, we finally find the nonlocal and covariant anomaly effective action as
\begin{align}
&\mathcal{S}_{\rm anom}^{}(g) =\frac {1}{8}\!\int \!d^4x\sqrt{-g_x}\, \left(E - \frac{2}{3}\sq R\right)_{\!x} \nonumber\\
&\int\! d^4x'\sqrt{-g_{x'}}\,D_4(x,x')\left[\frac{b'}{2}\, \left(E - \frac{2}{3}\sq R\right) \right]_{x'}.
\label{Snonl}
\end{align}
 \section{ Constraints on $V_E$ in $d$ dimensions for Lovelock gravity}
\label{heres}
The presence in 4d EGB of a topological term has important implications concerning the structure of such classical contributions. This point can be understood more clearly by discussing the role of the term $g_s(d)V_E$ in the context of the conformal anomaly effective action \cite{Coriano:2022ftl}. This action is naturally derived from a path integral, once we integrate out a conformal sector, and one can show that the counterterm sector - that in this case involves also the square of the Weyl tensor - separately satisfies anomalous CWIs. \\
 The derivation of such identities follows a direct pattern, that consists in writing down the conformal anomaly action in a background metric endowed with conformal Killing (CKVs) vectors.  In this section we illustrate the derivation of these constraints that define, in the context of the $4d$ GB theory, the application of this method. This point can be understood geometrically in the following way. \\
In a local free falling frame of a curved spacetime (i.e. in tangent space), we require that a certain action is endowed with a conformal symmetry, enlarging the usual local Poincar\`e symmetry of Einstein's theory.  Such conformal symmetry of each local frame can be gauged in the form of a general metric that allows CKVs. In the case of a conformal anomaly action, the contribution coming from $V_E$, as already mentioned, is paired with $V_{C^2}$. The latter is the only effective counterterm needed in order to remove the singularity of the quantum corrections at $d=4$. The CWIs of the complete effective action get splitted into three separate contributions: those derived from the renormalized quantum corrections and those associated with $V_E$ and $V_{C^2}$. 
All the three contributions satisfy separate conservation WIs and CWIs. Those corresponding to the finite renormalized quantum corrections, once we perform the flat spacetime limit, are ordinary, while the other two hierarchies, related to $V_E$ and $V_{C^2}$, are anomalous.  The constraints on these functionals come from their response once we perform a variation respect to the conformal factor $\phi$. \\
We detail the derivation of this point. \\
 We recall that the CKVs are solutions of the equation

\begin{equation}
\nabla_\mu \xi_\nu + \nabla_\nu \xi_\mu = \frac{2}{d} \nabla_\lambda \xi^\lambda g_{\mu \nu} .
\label{bb}
\end{equation}
To derive the CWI's in the flat limit, we need to require that the background metric allows CKVs that leave the action invariant. We start from the conservation of the conformal current 
\begin{equation}
	\int d^d x \sqrt{g}\, \nabla_\mu \left( \xi_\nu^{(K)} V_E^{\mu\nu}\right)=0,
\label{eew}
\end{equation}
and, analogously, we can write
\begin{equation}
	\int d^d x \sqrt{g}\, \nabla_\mu \left( \xi_\nu V_E^{\mu\nu\mu_1\nu_1}\right)=0.  \label{Ve2}
\end{equation}
The identity
\begin{align} \label{CKVconflat11}
&2 \delta_{\mu \nu} \delta^{\lambda \rho} \xi_\lambda \partial_\rho \phi = 
2 \delta_{\mu \nu} e^{2 \phi} g^{\lambda \rho} \xi_\lambda \partial_\rho \phi = \nonumber\\
&2 \delta_{\mu \nu} e^{2 \phi} \xi^\lambda \partial_\lambda \phi
\end{align} 
can be used in the covariant derivatives of \eqref{bb} to obtain 
\begin{align} \label{CKVconflat1}
&\partial_\mu \xi_\nu - 2 \xi_\mu \partial_\nu \phi + \partial_\nu \xi_\mu - 2\xi_\nu \partial_\mu \phi + 2 \delta_{\mu \nu} \delta^{\lambda \rho} \xi_\lambda \partial_\rho \phi=\nonumber\\
& \frac{2}{d} \partial_\lambda \xi^\lambda e^{2 \phi} \delta_{\mu \nu} + 2 \delta_{\mu \nu} e^{2 \phi} \xi^\lambda \partial_\lambda \phi. 
\end{align}
This can be written  in the form 
\begin{align} \label{CKVconflat12}
&\partial_\mu \xi_\nu - 2 \xi_\mu \partial_\nu \phi + \partial_\nu \xi_\mu - 2\xi_\nu \partial_\mu \phi = \nonumber\\
&e^{2 \phi}[ \partial_{\mu} (e^{- 2 \phi} \xi_\nu ) + \partial_{\nu} (e^{- 2 \phi} \xi_\mu ) ] =\nonumber\\
&e^{2 \phi} (\delta_{\nu \lambda} \partial_\mu + \delta_{\mu \lambda} \partial_\nu )\xi^\lambda ,
\end{align}
Substituting \eqref{CKVconflat11} and \eqref{CKVconflat12} in \eqref{CKVconflat1} we get
\eqref{bb} with the ordinary derivative replacing the covariant ones
\beq
\label{flatc}
	\partial_\mu \xi_\nu+\partial_\nu \xi_\mu=\frac{2}{d}\delta_{\mu\nu}\,\left(\partial\cdot \xi\right).
\eeq
Writing explicitly the action of the covariant derivative in \eqref{eew} and taking the flat limit, then we obtain the constraint
\begin{align}
	0=\int\,d^dx\,\bigg(\partial_\mu \xi_\nu\,V_{E}^{\mu\nu\mu_1\nu_1}+ \xi_\nu\,\partial_\mu\,V_E^{\mu\nu\mu_1\nu_1}\bigg).
\end{align}
We recall that $\xi_\nu$ satisfies the conformal Killing equation in flat space
and by using \eqref{Ve2}, the equation above can be re-written in the form
\begin{align}
	0=\int\,d^dx\,\bigg(\xi_\nu\,\partial_\mu V_E^{\mu\nu\mu_1\nu_1}+\frac{1}{d}\big(\partial\cdot \xi\big)\delta_{\mu\nu}\,V_E^{\mu\nu\mu_1\nu_1}\bigg).\label{newcons}
\end{align}
If we use in  this previous expression the conservation and the trace identities for $V_E^{\mu\nu\mu_1\nu_1}$, that are explicitly given by
\begin{align}
&\partial_\mu V_E^{\mu\nu\mu_1\nu_1}(x,x_1)=\bigg(\delta^{(\mu_1}_\mu\delta^{\nu_1)}_\lambda\partial^\nu\delta(x-x_1)-\nonumber \\
&2\delta^{\nu(\mu_1}\delta^{\nu_1)}_\mu\partial_\lambda\delta(x-x_1)\bigg)V_E^{\lambda\mu}(x),\label{consTT}\nonumber \\
&\delta_{\mu\nu}V_E^{\mu\nu\mu_1\nu_1}(x,x_1)=2(d-4)\big[\sqrt{-g(x)}E(x)\big]^{\mu_1\nu_1}(x_1)-\nonumber \\
&2\delta(x-x_1)V_E^{\mu_1\nu_1}(x)
\end{align}
together with the explicit expression of the Killing vector $\xi^{(C)}_\nu$ for the special conformal transformations, we derive the relation 
\begin{equation}
	\begin{split}
		\xi^{(C)\,\kappa}_\mu&=2x^\kappa\,x_\mu-x^2\delta^\kappa_\mu\\
		\partial\cdot  \xi^{(C)\,\kappa}&=2d\,x^\kappa
	\end{split}\label{spc}
\end{equation}
where $\kappa=1,\dots,d$. By using \eqref{spc} in the integral \eqref{newcons}, we derive the relation \begin{align}
	0=\int\,d^dx\,\bigg[\big(2x^\kappa\,x_\nu-x^2\delta^\kappa_\nu\big)\partial_\mu\,V_E^{\mu\nu\mu_1\nu_1}
\nonumber \\
	+2\,x^\kappa\,\delta_{\mu\nu}V_E^{\mu\nu\mu_1\nu_1}&\bigg]
\end{align}
that can be cast into the form
\begin{align}
&0=\int\,d^dx\,\bigg[\big(2x^\kappa\,x_\nu-x^2\delta^\kappa_\nu\big)\bigg(\delta^{(\mu_1}_\mu\delta^{\nu_1)}_\lambda\partial^\nu\delta(x-x_1)-\nonumber \\
& 2\delta^{\nu(\mu_1}\delta^{\nu_1)}_\mu\partial_\lambda\delta(x-x_1)\bigg)V_E^{\lambda\mu}(x)
-4\,x^\kappa\,\delta(x-x_1)V_E^{\mu_1\nu_1}(x)\bigg].
\end{align}
Integrating by parts, we obtain the expression
\begin{align}
&\left(2d\,x_1^\kappa+2x_1^\kappa\,x^{\mu}_1\frac{\partial}{\partial x_1^\mu}+x_1^2\frac{\partial}{\partial x_{1\kappa}}\right)V_E^{\mu_1\nu_1}+\nonumber\\
&2\bigg(x_{1\lambda}\,\delta^{\mu_1\kappa}-x_1^{\mu_1}\delta^\kappa_\lambda\bigg)V_E^{\lambda\nu_1}(x_1)\notag\\
&+2\bigg(x_{1\lambda}\,\delta^{\nu_1\kappa}-x_1^{\nu_1}\delta^\kappa_\lambda\bigg)V_E^{\mu_1\lambda}(x_1)=\nonumber\\
&4(d-4)\,\int dx\,x^\kappa\big[\sqrt{-g(x)}E(x)\big]^{\mu_1\nu_1}(x_1)
\end{align}
which is the special CWIs satisfied by $V_E^{\mu_1\nu_1}$. This relation is, in this case, trivially satisfied just because $V_E^{\mu_1\nu_1}$, in the flat limit, vanishes. 
Non trivial constraints are generated when we consider the contributions - starting  from the three graviton vertex onwards - and we have for $n\ge3$
\begin{align}
&\sum_{j=1}^n\biggl[2x_j^\kappa\biggl(d+x_j^\alpha\frac{\partial}{\partial x_j^\alpha}\biggl)-x_j^2\,\delta^{\kappa\alpha}\frac{\partial}{\partial x_j^\alpha}\biggl]\,\nonumber\\
&V_E^{\mu_1\nu_1\dots\mu_n\nu_n}(x_1,\dots,x_n)+\notag\\
&2\sum_{j=1}^n\biggl(\delta^{\kappa\mu_j}x_{j\,\alpha}-\delta^\kappa_\alpha x_j^{\mu_j}\biggl)\,\nonumber\\
&V_E^{\mu_1\nu_1\dots\nu_j\alpha\dots\mu_n\nu_n}(x_1,\dots,x_j,\dots,x_n)+\notag\\
&2\sum_{j=1}^n\biggl(\delta^{\kappa\nu_j}x_{j\,\alpha}-\delta^\kappa_\alpha x_j^{\nu_j}\biggl)\nonumber\\
&V_E^{\mu_1\nu_1\dots\mu_j\alpha\dots\mu_n\nu_n}(x_1,\dots,x_j,\dots,x_n)=\notag\\
&2^{n+1}(d-4)\,\int d^dx\,x^\kappa\biggl[\sqrt{-g(x)}\,\nonumber\\
&E(x)\biggl]^{\mu_1\nu_1\dots\mu_n\nu_n}(x_1,\dots,x_n).
\end{align}
The dilatation CWI is obtained by the choice of the CKV of the form
\begin{equation}
	\xi^{(D)}_\mu(x)=x_\mu,\qquad\partial\cdot \xi^{(D)}=d,
\end{equation}
and equation \eqref{newcons}, in the general case becomes
\begin{align}
&0=\int d^d x\bigg\{x_\mu\,\partial_\nu V_E^{\mu\nu\mu_1\nu_1\dots\mu_n\nu_n}(x,x_1,\dots,x_n)+\nonumber\\
&\delta_{\mu\nu}V_E^{\mu\nu\mu_1\nu_1\dots\mu_n\nu_n}(x,x_1,\dots,x_n)\label{anomDil}\bigg\}.
\end{align}
Taking into account the conservation and trace identities satisfied by $V_E^{\mu_1\nu_1\dots\mu_n\nu_n}$ we obtain the final expression
\begin{align}
	&\left(n\,d+\sum_{j=1}^n\,x_j^\alpha\frac{\partial}{\partial x_j^\alpha}\right)\,V_E^{\mu_1\nu_1\dots \mu_n\nu_n}(x_1,\dots,x_n)=\nonumber\\
	&2^n(d-4)\int d^dx \left[\sqrt{-g(x)}\,E(x)\right]^{\mu_1\nu_1\dots\mu_n\nu_n}(x_1,\dots,x_n)\label{DilAnom}.
\end{align}
This constraint is non trivial starting from $n=3$. 
In momentum space these equations are written as
\begin{align}
	&\left(d-\sum_{j=1}^{n-1}\,p_j^\alpha\frac{\partial}{\partial p_j^\alpha}\right)\,V_E^{\mu_1\nu_1\dots \mu_n\nu_n}(p_1,\dots,\bar{p}_n)=\nonumber\\
	&2^n(d-4)\left[\sqrt{-g}\,E\right]^{\mu_1\nu_1\dots\mu_n\nu_n}(p_1,\dots,\bar{p}_n),\label{DilAnomMom}
\end{align}
 and 
\begin{align}
	&\sum_{j=1}^{n-1}\left(p_j^\kappa\frac{\partial^2}{\partial p_j^\alpha \partial p_{j\alpha}}-2p_j^\alpha\frac{\partial}{\partial p_j^\alpha\partial p_{j\kappa}}\right)V_E^{\mu_1\nu_1\dots \mu_n\nu_n}(p_1,\dots,\bar{p}_n)\notag\\
	&+2\sum_{j=1}^{n-1}\left(\delta^{\kappa\mu_j}\frac{\partial}{\partial p_{j\,\alpha}}-\delta^\kappa_\alpha \frac{\partial}{\partial p_j^{\mu_j}}\right)\nonumber\\
	&V_E^{\mu_1\nu_1\dots\nu_j\alpha\dots\mu_n\nu_n}(p_a,\dots,p_j,\dots\bar{p}_n)\notag\\
	&+2\sum_{j=1}^{n-1}\left(\delta^{\kappa\nu_j}\frac{\partial}{\partial p_{j\,\alpha}}-\delta^\kappa_\alpha \frac{\partial}{\partial p_j^{\nu_j}}\right)\nonumber\\
	&V_E^{\mu_1\nu_1\dots\mu_j\alpha\dots\mu_n\nu_n}(p_a,\dots,p_j,\dots\bar{p}_n)\notag\\
	&=-2^{n+1}(d-4)\nonumber\\
	&\bigg[\frac{\partial}{\partial p_{n\kappa}}\bigg(\,\left[\sqrt{-g}\,E\right]^{\mu_1\nu_1\dots\mu_n\nu_n}(p_1,\dots,p_n)\bigg)\bigg]_{p_n=\bar{p}_n},
\label{fff}
\end{align}
where $\bar{p}_n=-\sum_{i=1}^{n-1}p_i$ for the conservation of the total momentum. These constraints are directly satisfied in $d\neq 4 $ dimensions and are therefore typical of Lovelock's theories of gravity in generic dimensions. Notice that the equations above are modified by an overall dimensionful factor both on their lhs and rhs, for dimensional reasons, but leaving \eqref{fff} unaltered. In the $d\to 4$ limit the analysis of these equations requires an accurate study of the degeneracy of such such structures, similarly to the one performed in \cite{Coriano:2022jkn}.  \\
In the context of anomaly actions, their reduction to $d\to 4$ requires the inclusion of a Weyl invariant sector, provided by the finite quantum corrections, missing here.
\subsection{Conservation identity at $d=4$ and trace identities at $d\neq 4$}
The same vertex satisfies a hierarchy of of conservation identity at $d=4$ in flat space, starting from a  curved background. Trace Ward identities, instead, are valid for this vertex at $d\neq 4$, as we are going to show next. Indeed, from 
 \begin{equation}
\label{vat}
\nabla_\mu V_E^{\mu\nu}(x)_g=0, 
\end{equation}
 expanding the covariant derivative, we obtain the relation
\begin{align}
&\partial_{\nu_1}V_E^{\mu_1\nu_1 \mu_2\nu_2 \mu_3\nu_3 \mu_4\nu_4}(x_1,x_2,x_3,x_4)=\notag\\
&-\biggl[2\left(\frac{\delta\Gamma^{\mu_1}_{\lambda\nu_1}(x_1)}{\delta  g_{\mu_2\nu_2}(x_2)}\right)_{g=\delta}\nonumber\\
&V_E^{\lambda\nu_1 \mu_2\nu_2 \mu_3\nu_3 \mu_4\nu_4}(x_1,x_2,x_3,x_4)+(23)+(24)\biggl]\notag\\
& -\biggl[4\left(\frac{\delta^2\Gamma^{\mu_1}_{\lambda\nu_1}(x_1)}{\delta  g_{\mu_2\nu_2}(x_2)\delta  g_{\mu_3\nu_3}(x_3)}\right)_{g=\delta}V_E^{\lambda\nu_1}(x_1, x_4)+\nonumber\\
&(24)+(34)\biggl],\label{transverseX}
\end{align}
where 
\begin{align}
&\biggl(\frac{\delta\Gamma^{\mu_1}_{\lambda\nu_1}(x_1)}{\delta  g_{\mu_i\nu_i}(x_i)}\biggl)_{g=\delta}=\frac{1}{2}\biggl(\delta^{\mu_1(\mu_i}\delta^{\nu_i)}_{\nu_1}\,\partial_\lambda\delta_{x_1x_i}+\nonumber\\
&\delta^{\mu_1(\mu_i}\delta^{\nu_i)}_{\lambda}\,\partial_{\nu_1}\delta_{x_1x_i}-\delta^{(\mu_i}_\lambda\delta^{\nu_i)}_{\nu_1}\,\partial^{\mu_1}\delta_{x_1x_i}\biggl)\nonumber\\[1.5ex]
%%%%%%%%%%%%%%%%%%%%%%%%%%%%%%%%%%%%%%%%%%%%%%%%%%%%%%%%%%%%%%%%%%%%%%%%%%%%%%%%%%%%%%%%%%%%%%
&\left(\frac{\delta^2\Gamma^{\mu_1}_{\lambda\nu_1}(x_1)}{\delta  g_{\mu_i\nu_i}(x_i)\delta  g_{\mu_j\nu_j}(x_j)}\right)_{g=\delta}=\notag\\
&-\frac{\delta_{x_1x_i}}{2}\delta^{\mu_1(\mu_i}\delta^{\nu_i)\epsilon}\biggl(\delta^{(\mu_j}_\epsilon\delta^{\nu_j)}_{\nu_1}\,\partial_\lambda\delta_{x_1x_j}+\nonumber\\
&\delta^{(\mu_j}_\epsilon\delta^{\nu_j)}_{\lambda}\,\partial_{\nu_1}\delta_{x_1x_j}-\delta^{(\mu_j}_\lambda\delta^{\nu_j)}_{\nu_1}\,\partial_\epsilon\delta_{x_1x_j}\biggl)+(ij),
%%%%%%%%%%%%%%%%%%%%%%%%%%%%%%%%%%%%%%%%%%%%%%%%%%%%%%%%%%%%%%%%%%%%%%%%%%%%%%%%%%%%%%%%%%%%%
\end{align}
that in momentum space becomes 
\begin{align}
&p_{1\,\nu_1}\,V_{E}^{\m_1\n_1\m_2\n_2\m_3\n_3\mu_4\nu_4}(p_1,p_2,p_3,\bar{p}_4)=\notag\\
&=\Big[4\, \mathcal{B}^{\mu_1\hspace{0.4cm}\mu_2\nu_2\mu_3\nu_3}_{\hspace{0.3cm}\lambda\nu_1}(p_2,p_3)V_{E}^{\lambda\n_1\mu_4\nu_4}(p_1+p_2+p_3,\bar{p}_4)\nonumber\\
&+(34)+ (24)\Big]\notag\\
&+\Big[2 \, \mathcal{C}^{\mu_1\hspace{0.4cm}\mu_2\nu_2}_{\hspace{0.3cm}\lambda\nu_1}(p_2)V_{E}^{\lambda\n_1\m_3\n_3\mu_4\nu_4}(p_1+p_2,p_3,\bar{p}_4)+\nonumber\\
&(2 3)+(24)\Big],
\end{align} 
where 	
\begin{align}
\label{BB}
\mathcal{B}^{\mu_1\hspace{0.4cm}\mu_2\nu_2\mu_3\nu_3}_{\hspace{0.3cm}\lambda\nu_1}(p_2,p_3)\equiv -\frac{1}{2}\delta^{\mu_1(\mu_2}{\delta^{\nu_2)\epsilon}}&\times \nonumber \\
 \times\left(\delta_\epsilon^{(\mu_3}\delta^{\nu_3)}_{\nu_1}\,p_{3\,\lambda}+\delta_\epsilon^{(\mu_3}\delta^{\nu_3)}_{\lambda}\,p_{3\,\nu_1}-\delta_\lambda^{(\mu_3}\delta^{\nu_3)}_{\nu_1}\,p_{3\,\epsilon}\right)+(23),&\nonumber\\
\end{align}
and
\begin{align}
\mathcal{C}^{\mu_1\hspace{0.4cm}\mu_2\nu_2}_{\hspace{0.3cm}\lambda\nu_1}(p_2)&\equiv \frac{1}{2}\left(\delta^{\mu_1(\mu_2}\delta^{\nu_2)}_{\nu_1}\,p_{2\,\lambda}+\delta^{\mu_1(\mu_2}\delta^{\nu_2)}_{\lambda}p_{2\,\nu_1}\right.  &\nonumber \\
&\left. -\delta^{(\mu_2}_{\lambda}\delta^{\nu_2)}_{\nu_1}p_2^{\mu_1}\right),
\end{align}
correspond, respectively, to the second and first functional derivatives of the Christoffel connection.

 A similar analysis can be performed for a trace identity. In this case, we simply functionally differentiate the  anomalous Weyl variation \eqref{epx} multiple times and transform the expression to momentum space, obtaining  
 \begin{align}
\label{mom}
&\delta_{\mu_1\nu_1}\,V_{E}^{\m_1\n_1\dots\mu_n\nu_n}(p_1,\dots,p_n)=\nonumber\\
&2^{n-1}(d-4)\,\left[\sqrt{-g}E\right]^{\mu_2\nu_2\dots\mu_n\nu_n}(p_2,\dots,p_n)\notag\\
&-2\bigg[V_{E}^{\m_2\n_2\dots\mu_n\nu_n}(p_1+p_2,p_3,\dots,p_n)+\nonumber\\&
V_{E}^{\m_2\n_2\dots\mu_n\nu_n}(p_2,p_1+p_3,\dots,p_n)+\dots+\nonumber\\
&V_{E}^{\m_2\n_2\dots\mu_n\nu_n}(p_2,p_3,\dots,p_1+p_n)\bigg].
\end{align}
This constraint is satisfied by all the vertices extracted from the $E_4$ term, and remain valid 
once this contribution is included in a Lovelock action when $d\neq 4$.

\section{Classical constraints on the equations of motion in the local action} 
\label{cla}
The constraints derived in the previous sections, as already mentioned, are obtained by performing the flat spacetime limit of the metric variations, without resorting to a conformal decomposition of the metric itself. \\
More general constraints are obtained if we, instead, perform a conformal decomposition and vary 
the fiducial metric and the conformal factor independently. The separation is consistent with the fact that the subtractions included in the definition of the WZ action introduce a conformal scale. This separation is scale invariant, in the sense that the resulting action is free of any dimensionful constant. These types of actions are typical of dilaton gravities and can be modified by the addition of extra scale-invariant potentials.    \\
In this section we investigate the consistency of the equations of motion, discussing their conservation in the local version of the theory. The same consistency will be missing once we move to the nonlocal theory, obtained by eliminating the dilaton, using the Riegert decomposition \eqref{onshell}. \\
As we are going to illustrate in a final section, in that case we need to amend the action by Weyl invariant contributions, that are necessary in order to derive the exact 
expression  of the hierarchy. This, in principle, requires an analysis of the $4T$ correlator, correcting the 
predictions derived from the nonlocal actions, with extra Weyl-invariant terms. Such terms can be identified by an approach that has been already discussed for a simpler correlator, the $TTJJ$ \cite{Coriano:2022jkn}. While this is an important point that we hope to discuss elsewhere, it is possible to obtain the correct hierarchies satisfied by the 4-point vertices of a 
4d GB theory by resorting directly to a previous analysis of the counterterms of the same vertex. This study has been presented in \cite{Coriano:2021nvn}.  \\
\subsection{The dilaton gravity action}
Coming to the local dilaton-gravity form of the limiting theory, the two fields can be treated as independent, but their equations will be linked by the constraints coming from the anomalous variation of the Euler Poincar\`e density.
For this reason, \eqref{rdef} defines a dilaton gravity theory in which the trace of the equations of motion of the gravity 
metric $\bar g$ and that of the conformal factor are related in the form 
\beqa
\label{fin}
\left(2 \bar{g}_{\mu\nu}\frac{\delta}{\delta \bar g_{\mu\nu}}-\frac{\delta}{\delta\phi}\right)\mathcal{S}^{(WZ)}_{GB}=-\alpha\sqrt{\bar g}\bar{E}.
\eeqa
The derivation of this relation is discussed in \cite{Coriano:2022ftl} for conformal anomaly actions. Note that the regulated action \eqref{WZ} is, separately, a functional of 
$\bar{g}$ and $\phi$ and one can use the relations
\beq
2 g_{\mu\nu}\frac{\delta }{\delta g_{\mu\nu}}V_E(g)=\frac{\delta}{\delta\phi}V_E(g)=\epsilon \sqrt{g} E,
\label{wwi}
\eeq
\beq
2 \bar{g}_{\mu\nu}\frac{\delta }{\delta \bar{g}_{\mu\nu}}V_E(\bar g)=2 {g}_{\mu\nu}\frac{\delta }{\delta {g}_{\mu\nu}}V_E(\bar g), 
=\epsilon \sqrt{\bar g}\bar E \eeq
(using $\bar{g}_{\mu\nu}=g_{\mu\nu}e^{-2 \phi}$) and 
\beq
\frac{\delta}{\delta \phi}V_E(\bar g)=0
\eeq
to obtain \eqref{fin}.

It is convenient to define the two tensors
\beq
T^{\mu\nu}=\frac{2}{\sqrt{\bar g}}\frac{\delta \sm_{WZ}}{\delta\bar g_{\mu\nu}}
\eeq   
and
\beq
T_\phi=\frac{1}{\sqrt{\bar g}}\frac{\delta \sm_{WZ}}{\delta\phi}.
\eeq

The relation can also be obtained by a direct computation using \eqref{rdef} 
\begin{align}
&\sqrt{\bar g}T_\phi=\sqrt{g}\biggl(\bar E + 8 \bar G_{\mu\nu}\bar \nabla_\mu\phi\bar \nabla_{\nu}\phi \nonumber\\
&+ 8 \bar \Box \phi \bar\nabla_\mu\phi \bar \nabla^\mu\phi +16 \bar \nabla_\lambda\bar \nabla_\mu\phi \bar \nabla^\lambda \phi \bar \nabla^\mu\phi \nonumber \\
& \qquad\qquad  -8 \bar R_{\mu\nu}\bar \nabla^\mu\phi\bar \nabla^\nu\phi + 8 (\bar \Box\phi)^2 -\nonumber\\
&8\bar \nabla_\mu\bar \nabla_\nu\phi \bar \nabla^\mu\bar \nabla^\nu\phi\biggl)\equiv  \sqrt{\bar g} \bar g_{\mu\nu}T^{\mu\nu}
\end{align}
 where the last equality follows from the trace of \eqref{ep3}. Rescalings in the conformal decomposition are typically of the form 
 \bea
 &R_{\mu \nu \rho \sigma}^2 =e^{-4\phi} \biggl(\bar R_{\mu \nu \rho \sigma}^2 - 8 \bar R^{\mu \nu} \bar \Delta_{\mu \nu} - \nonumber\\
&4 \bar R\bnabla_\lambda \phi \bnabla^\lambda \phi + 4(d-2)\bar \Delta_{\mu \nu}^2 +  \nonumber\\
&4\bar \Delta ^2+ 8(d-1)\bar \Delta \bnabla_\lambda \phi \bnabla^\lambda \phi  +\nonumber\\ 
&2d(d-1)(\bnabla_\lambda \phi \bnabla^\lambda \phi)^2 \biggl) 
\eea
and similar ones. They can be found in \cite{Coriano:2022ftl}.
The action is diffeomorphism invariant since $\phi$ transforms as a scalar under changes of coordinates. We can use this invariance to derive the equation satisfied by the stress energy tensor, by varying the action with respect to the fiducial metric. The Lie derivatives for the scalar field $\phi$ and the fiducial metric $\bar g_{\mu\nu}$ are
\begin{equation}
    \begin{cases}
     \delta_\xi \phi=\xi^\lambda\bar\nabla_\lambda\phi\\
     \de_\xi\bar g_{\mu\nu}=\bar g_{\mu\lambda}\bar \nabla_\nu \xi^\lambda+\bar g_{\nu\lambda}\bar \nabla_\mu\xi^\lambda
         \end{cases}\,.
\end{equation}
giving the variation
\begin{align}
&\delta \sm_{WZ}=\int d^d x \left ( \de_\xi\phi \frac{\delta}{\delta \phi}+\delta_\xi\bar g_{\mu\nu}\frac{\delta}{\delta\bar g_{\mu\nu}}   \right )\sm_{WZ} \notag\\
=&\int d^d x \left ( \xi^\lambda\bar\nabla_\lambda\phi\frac{\delta}{\delta\phi}+(\bar g_{\mu\lambda}\bar \nabla_\nu \xi^\lambda+\bar g_{\nu\lambda}\bar \nabla_\mu\xi^\lambda)\frac{\delta}{\delta\bar g_{\mu\nu}}   \right)\sm_{WZ}\notag\\
=&\int d^d x\xi^\lambda \left ( \bar\nabla_\lambda\phi\frac{\delta}{\delta\phi}-2\bar g_{\mu\lambda}\bar \nabla_\nu\frac{\delta}{\delta\bar g_{\mu\nu}}  \right )\sm_{WZ}.
\end{align}
The condition to be imposed to get the invariance under diffeomorphism for a generic functional such as $\sm_{WZ}$ is 
\beq
\label{diff inv}
\left ( \bar\nabla_\lambda\phi\frac{\delta}{\delta\phi}-2\bar g_{\mu\lambda}\bar \nabla_\nu\frac{\delta}{\delta\bar g_{\mu\nu}}  \right )\sm_{WZ}=0.
\eeq
From now to the rest we will omit the "bar" above all the tensor, derivatives ecc., assuming that they are all evaluated respect to the fiducial metric $\bar{g}$, treated as an independent field. The first variation gives
\begin{align}
\label{first term}
&\nabla_\lambda\phi\frac{\delta }{\delta\phi}\mathcal{S}_{WZ}=\nabla_\lambda\phi\biggl [ E+ 8G^{\mu\nu}\nabla_\mu\nabla_\nu \phi+\nonumber\\
&8\Box\phi\nabla_\mu\phi\nabla^\mu\phi+16\nabla_\lambda\nabla_\mu\phi\nabla^\lambda\phi\nabla^\mu\phi\notag \\
&-8R_{\mu\nu}\nabla^\mu\phi\nabla^\nu\phi+8(\Box\phi)^2-8\nabla_\mu\nabla_\mu\phi\nabla^\mu\nabla^\nu\phi\biggl].
\end{align}
The second variation is given by
\begin{align}
\label{second term}
&-2 g_{\mu\lambda} \nabla_\nu\frac{\delta}{\delta g_{\mu\nu}}\mathcal{S}_{WZ}=\nonumber\\
&-2 g_{\mu\lambda} \nabla_\nu\int d^4x \rg \frac{\delta}{\delta g_{\mu\nu}}(\phi E)\notag\\
&-\nabla_\lambda\phi\biggl [ 8G^{\mu\nu}\nabla_\mu\nabla_\nu \phi+8\Box\phi\nabla_\mu\phi\nabla^\mu\phi+\nonumber\\
&16\nabla_\lambda\nabla_\mu\phi\nabla^\lambda\phi\nabla^\mu\phi -8R_{\mu\nu}\nabla^\mu\phi\nabla^\nu\phi\notag\\
&+8(\Box\phi)^2-8\nabla_\mu\nabla_\mu\phi\nabla^\mu\nabla^\nu\phi\biggl]+\nonumber\\ 
&8R_{\lambda\mu\nu\rho}\nabla^\nu\phi(\nabla^\mu\nabla^\rho\phi-\nabla^\rho\nabla^\mu\phi).
\end{align}
Adding \eqref{first term} and \eqref{second term}, recalling that $[\nabla_\mu,\nabla_\nu]\phi=0$, we finally get
\begin{align}
&\left ( \nabla_\lambda\phi\frac{\delta}{\delta\phi}-2 g_{\mu\lambda} \nabla_\nu\frac{\delta}{\delta g_{\mu\nu}}  \right )\mathcal{S}_{WZ}=\nabla_\lambda\phi E-\nonumber\\
&2 g_{\mu\lambda} \nabla_\nu\int d^4x \rg \frac{\delta}{\delta g_{\mu\nu}}(\phi E).
\end{align}
The explicit form of the second term in the equation above is expressed by
\begin{align}
&-2 g_{\mu\lambda} \nabla_\nu\int d^4x \rg \frac{\delta}{\delta g_{\mu\nu}}(\phi E)=-\nabla_\lambda\phi E-\nonumber\\
&4R^{\nu\mu\alpha\beta}\phi\nabla_\mu R_{\lambda\nu\alpha\beta}-2R^{\nu\mu\alpha\beta}\phi\nabla_\lambda R_{\nu\mu\alpha\beta}\notag\\
&+4\nabla^\mu\nabla^\nu\phi(\nabla_\nu R_{\lambda\mu}- \nabla_\mu R_{\lambda\nu})-\nonumber\\
&4(R_{\lambda\mu\nu\alpha}+R_{\lambda\nu\mu\alpha})\nabla^\alpha\nabla^\mu\nabla^\nu\phi+4R_{\nu\mu\alpha\beta}R_{\lambda}{}^{\mu\alpha\beta}\nabla^\nu\phi\notag\\
&+2\phi[\nabla_\lambda,\nabla_\nu]\nabla^\nu R+2R_{\lambda\nu}\phi\nabla^\nu R.
\end{align}
The simplification of this expression requires some intermediate steps.
 A lengthy but direct computation shows that the Bianchi identity 
\begin{equation}
R^\mu_{\ \nu \rho \sigma} + R^\mu_{\ \rho \sigma \nu} + R^\mu_{\ \sigma \nu \rho} = 0 
\end{equation}
is still satisfied for a conformal decomposition.
Thanks to the above equation, using the symmetry of the  Riemann tensor and 
\beq
[\nabla_\nu,\nabla_\mu]V^\rho=R^\rho{}_{\sigma\mu\nu}V^{\sigma},
\eeq
it is quite easy to show \eqref{diff inv}. A similar computation can be performed to derive \eqref{fin} from 
\eqref{rdef}.\\
It is quite obvious that the singular procedure that takes to a regulated $4d$ EGB action is consistent.  
This analysis becomes rather nontrivial as the dilaton is removed from the spectrum. As already mentioned, in that case the nonlocal action needs to be amended by extra terms. This will occur at the level of the classical 4-graviton vertex. The 3-graviton vertex, instead, can be handled directly with the nonlocal action \eqref{Snonl}. Results of this analysis are given below.
\section{3- and 4-wave interaction in the nonlocal $4d$ GB theory}
The nonlocal structure of the $4d$ EGB theory results from an iterative solution of the equations of motion in which the dilaton is expressed in terms of the full original metric $g$, as shown in \cite{Coriano:2017mux}.  
One can rewrite the nonlocal  action in the form 
\bea
\label{loc}
& \cS_{\rm anom}(g,\vf) \equiv -\sdfrac{1}{2} \int d^4x\,\sqrt{-g}\, \Big[ (\sq \vf)^2 - \nonumber\\
&2 \big(R^{\m\n} - \tfrac{1}{3} R g^{\m\n}\big)
(\na_\m\vf)(\na_\n \vf)\Big]\nn
& +\, \sdfrac{1}{2}\,\int d^4x\,\sqrt{-g}\  \Big[\big(E - \tfrac{2}{3}\sq R\big)  \Big]\,\vf,
\label{Sanom}
\eea
that can be varied with respect to $\phi$, giving
\be
\sqrt{-g}\,\D_4\, \vf = \sqrt{-g}\left[\sdfrac{E}{2}- \sdfrac{\!\sq R\!}{3} \right] \label{phieom}.
\ee
Three-wave interactions can be derived by expanding perturbatively in the metric fluctuations in the form 
\bes
\bea
&g_{\m\n} = g_{\m\n}^{(0)} + g_{\m\n}^{(1)} + g_{\m\n}^{(2)} + \dots \equiv\nonumber\\
& \eta_{\m\n} + h_{\m\n} + h_{\m\n}^{(2)} + \dots\\
&\vf = \vf^{(0)} +  \vf^{(1)} +  \vf^{(2)}  + \dots
\eea
\ees
The expansion above should be interpreted as a collection of terms generated by setting 
\beq
g_{\m\n} = \delta_{\mu\nu} + \kappa h_{\m\n} 
\eeq
having reinstated the coupling expansion $\kappa$, with $h$ of mass-dimension one,  
and collecting all the higher order terms in the functional expansion of \eqref{loc} of the order $h^2$, $h^3$ and so on. A similar expansion holds for $\vf$ if we redefine $ \vf^{(1)}=\kappa \bar\vf^{(1)},  \vf^{(2)}=\kappa^2 \bar \vf^{(2)}$ and so on. 
At cubic level the vertex is given by 

\begin{align}
&\cS_{\rm nonl}^{(3)}\! =\!- \sdfrac{b'}{18}\! \int\! d^4x \left\{\!R^{(1)}\!\sdfrac{1}{\sqb} \big(\sqrt{-g} \sq^2\big)^{\!(1)}\sdfrac{1}{\sqb} R^{(1)}\! \right\}+\nonumber\\
& \sdfrac{b'}{9}\! \int\! d^4x \left\{\!\pa_{\m} R^{(1)}\!\sdfrac{1}{\sqb} \! \left(\!R^{(1)\m\n}\! - \!\sdfrac{1}{3} \eta^{\m\n} R^{(1)}\!\right)\!\sdfrac{1}{\sqb}\pa_{\n} R^{(1)}\!\right\} \nonumber\\
& - \sdfrac{1}{6}\! \int\! d^4x \left(b'\, E^{\!(2)} \right)\sdfrac{1}{\sqb}R^{(1)}+\nonumber\\
& \sdfrac{b'}{9} \!\int\! d^4x\,  R^{(1)}  \sdfrac{1}{\sqb} \left(\sqrt{-g}\sq\right)^{\!(1)}R^{(1)}
+ \sdfrac{b'}{9} \! \int\! d^4x\, R^{\!(2)}R^{(1)}
\label{Sanom3b}
\end{align}

where the suffixes $(1)$, $(2)$ denote the order of the expansion in the fluctuations around flat space $(g_{\mu\nu}=\delta_{\mu\nu} + h_{\mu\nu})$.
 From \eqref{Sanom3b} we can extract the expressions of the classical 3-wave gravitational interactions in this effective theory by differentiating three times with respect to $h_{\mu\nu}$.

\subsection{3-wave interactions in momentum space} 
The GB term in the equations of motion induces interactions of higher orders exhibiting specific features, that we are going to identify in this and in the next section. Obviously, cubic and quartic interactions in the nonlocal $4d $ EGB theory \eqref{Sanom3b} share close similarities with those identified in the nonlocal conformal anomaly 
action. At cubic level, the most convenient way to organize such contributions is to transform 
the expressions to momentum space. For this purpose,  we define
\beq
\int d^4x \, e^{-ip\cdot x} \, R_{\m\a\n\bet}^{(1)}(x) \equiv \big[R_{\m\a\n\bet}^{(1)}\big]^{\m_1\n_1}(p)\, \tilde h_{\m_1\n_1}(p)
\label{RieFour}
\eeq
for the linear expansion of the Ricci tensor
\be
R_{\m\a\n\bet}^{(1)} = \sdfrac{1}{2}\, \Big\{\!- \pa_\a\pa_\bet h_{\m\n}- \pa_\m\pa_\n h_{\a\bet} + \pa_\a\pa_\n h_{\bet\m}
+ \pa_\bet\pa_\m h_{\a\n}\Big\},
\ee
which in momentum space becomes
\begin{align}
&\big[R_{\m\a\n\b}^{(1)}\big]^{\m_1\n_1}(p) = \sdfrac{1}{2}\, \Big\{\d^{(\m_1}_\a\, \d^{\n_1\!)\hspace{-4pt}}\,_\b\,p_\m\, p_\n
+ \d^{(\m_1}_\m\, \d^{\n_1)}_\n\,p_\a\, p_\b  -\nonumber\\
& \d_\b\,^{\hspace{-2pt}(\m_1}\, \d^{\n_1)}_\m\,p_\a \,p_\n - \d^{(\m_1}_\a\, \d^{\n_1)}_\n\,p_\b \,p_\m  \Big\}.
\label{Riemom}
\end{align}
We will be needing the identities
\begin{align}
&\big[R_{\m\a\n\b}^{(1)}R^{(1)\m\a \n\b}\big]^{\m_1\n_1\m_2\n_2} (p_1, p_2) \equiv\nonumber\\
&\big[R_{\m\a\n\b}^{(1)}\big]^{\m_1\n_1} (p_1) \big[R^{(1)\m\a \n\b}\big]^{\m_2\n_2}(p_2) =\nonumber\\
& (p_1 \cdot p_2)^2\, \eta^{\m_1(\m_2}\eta^{\n_2)\n_1}
 - 2\, (p_1\cdot p_2)\, p_1\,^{\hspace{-4pt}(\m_2}\eta^{\n_2)(\n_1}p_2\,^{{\hspace{-2pt}}\m_1)}+\nonumber\\
 & p_1^{\m_2}\,p_1^{\n_2}\,p_2^{\m_1}\,p_2^{\n_1}
\label{Riemsq}
\end{align}
and
\vspace{-3mm}
\begin{align}
&\big[R_{\m\n}^{(1)}R^{(1)\m\n}\big]^{\m_1\n_1\m_2\n_2} (p_1, p_2) \equiv
\big[R_{\m\n}^{(1)}\big]^{\m_1\n_1} (p_1) \big[R^{(1)\m\n}\big]^{\m_2\n_2}(p_2) =\nonumber\\
&\sdfrac{1}{4}\, p_1^2 \, \Big(p_2^{\m_1}\,  p_2^{\n_1}\, \eta^{\m_2\n_2}  -  2\, p_2\,^{\hspace{-4pt}(\m_1}\eta^{\n_1)(\n_2}  p_2\,^{\hspace{-2.5pt}\m_2)}\Big)+\nonumber\\
& \sdfrac{1}{4}\, p_2^2\, \Big(p_1^{\m_2}\,  p_1^{\n_2}\, \eta^{\m_1\n_1} -  2\,p_1\,^{\hspace{-4pt}(\m_1}\eta^{\n_1)(\n_2}  p_1\,^{\hspace{-2.5pt}\m_2)}\Big)+\nonumber\\
& \sdfrac{1}{4}\, p_1^2\ p_2^2\, \eta^{\m_1(\m_2}\eta^{\n_2)\n_1}+\nonumber\\
& \sdfrac{1}{4}\, (p_1\cdot p_2)^2\, \eta^{\m_1\n_1}\eta^{\m_2\n_2} + \sdfrac{1}{2} \, p_1^{(\m_1}\,p_2^{\n_1)}\,p_1^{(\m_2}\,p_2^{\n_2)}+\nonumber\\
&  \sdfrac{1}{2} \, (p_1\cdot p_2)\,
\Big( p_1\,^{\hspace{-4pt}(\m_1}\, \eta^{\n_1)(\n_2}  p_2\,^{\hspace{-2.5pt}\m_2)}
-\eta^{\m_1\n_1}  \, p_1^{(\m_2}\,  p_2^{\n_2)} -\nonumber\\
&\eta^{\m_2\n_2}  \, p_1^{(\m_1}\,  p_2^{\n_1)}\Big)\,.
\label{Riccsq}
\end{align}
We also use the relation
\begin{align}
&\big[(R^{(1)})^2\big]^{\m_1\n_1\m_2\n_2} (p_1, p_2) \equiv\nonumber\\
&\big[R^{(1)}\big]^{\m_1\n_1} (p_1) \big[R^{(1)}\big]^{\m_2\n_2}(p_2)=\nonumber\\& p_1^2\, p_2^2 \, \pi^{\m_1\n_1}(p_1)\, \pi^{\m_2\n_2}(p_2),
\label{Riccscalsq}
\end{align}
and re-express the third order classical vertices of the $GB$ action and its contribution \eqref{Sanom3b} to the three-point correlator in momentum space, in the form
\begin{align}
&S_3^{\m_1\nu_1\m_2\nu_2\m_3\nu_3}(p_1,p_2,p_3)=\sdfrac{8}{3} \alpha\Big\{\pi^{\m_1\nu_1}(p_1)\,\nonumber\\
&\left[ E^{(2)}\right]^{\m_2\nu_2\m_3\nu_3}(p_2,p_3)+(\text{cyclic})\Big\}\nonumber\\
&\qquad -\sdfrac{16\alpha}{9}\Big\{ \pi^{\m_1\nu_1}(p_1)\,Q^{\m_2\nu_2}(p_1,p_2,p_3)\, \pi^{\m_3\nu_3}(p_3)\nonumber\\
&+(\text{cyclic})\Big\} +\sdfrac{16\alpha}{27\,}\, \pi^{\m_1\nu_1}(p_1)\,\pi^{\m_2\nu_2}(p_2)\,\pi^{\m_3\nu_3}(p_3)\,\nonumber\\
&\Big\{p_3^2\, p_1\cdot p_2+(\text{cyclic})\Big\},
 \label{AS3}
\end{align}
where we have defined
\begin{equation}
	\pi^{\mu\nu}_{}(p)\equiv\delta^{\mu\nu}-\frac{p^{\mu}p^{\nu}}{p^2},
\end{equation}
and
\begin{align}
&Q^{\m_2\nu_2}(p_1,p_2,p_3) \equiv p_{1\m}\, [R^{\m\nu}]^{\m_2\nu_2}(p_2)\,p_{3\n}= \nonumber\\
& \sdfrac{1}{2} \,\Big\{(p_1\cdot p_2)(p_2\cdot p_3)\, \delta^{\m_2\n_2} 
+ p_2^2 \ p_1^{(\m_2}\, p_3^{\n_2)} -\nonumber\\
& (p_2\cdot p_3) \, p_1^{(\m_2}\, p_2^{\n_2)} - (p_1\cdot p_2) \, p_2^{(\m_2}\, p_3^{\n_2)}\Big\}.
\end{align}
We have defined with 
\begin{align}
&\big[E^{(2)}\big]^{\m_i\nu_i\m_j\nu_j} =\big[R_{\m\a\n\b}^{(1)}R^{(1)\m\a \n\b}\big]^{\m_i\nu_i\m_j\nu_j}\nonumber\\
&-4\,\big[R_{\m\n}^{(1)}R^{(1)\m\n}\big]^{\m_i\nu_i\m_j\nu_j}+\big[ \big(R^{(1)}\big)^2\big]^{\m_i\nu_i\m_j\nu_j}
\end{align}
the second functional derivative of the topological density in flat space, after Fourier transform. One can prove the identity
\begin{align}
&\delta_{\a_1\b_1}\, S_3^{\a_1\b_1\m_2\n_2\m_3\n_3}(p_1,p_2,p_3)\Big\vert_{p_3 = -(p_1 + p_2)} =\nonumber\\
&\,8\alpha\, \big[E^{(2)}\big]^{\m_2\n_2\m_3\n_3} (p_2, p_3), 
 \label{traceS3}
\end{align}
together with the conservation identities
\bes
\bea
&&p_{2 \m_2} \, Q^{\m_2\n_2} (p_1, p_2, p_3) =  0\\
&&p_{2 \m_2} \, \big[E^{(2)}\big]^{\m_2\n_2\m_3\n_3} (p_2, p_3) = 0.
\eea
\ees
Double tracing \eqref{AS3} of the nonlocal theory one obtains 
\begin{align}
&\delta_{\a_1\b_1}\delta_{\a_3\b_3}\,S_3^{\a_1\b_1\m_2\n_2\a_3\b_3}(p_1,p_2,p_3)\big\vert_{p_3 = -(p_1 + p_2)} =\nonumber\\
& 8\alpha \, \delta_{\a_3\b_3}\big[E^{(2)}\big]^{\m_2\n_2\a_3\b_3} (p_2, p_3)  \nonumber\\
& \qquad = 16\alpha\, Q^{\m_2\n_2}(p_1,p_2,p_3)\big\vert_{p_3 = -(p_1 + p_2)} \, + \nonumber\\ 
&\ 8\alpha\,p_2^2\, \left( p_1^2  + p_1\cdot p_2\right)\pi^{\m_2\n_2}(p_2). 
\label{dbltrace3}
\end{align}
\begin{figure}
\label{ff2}
\center
\includegraphics[scale=0.45]{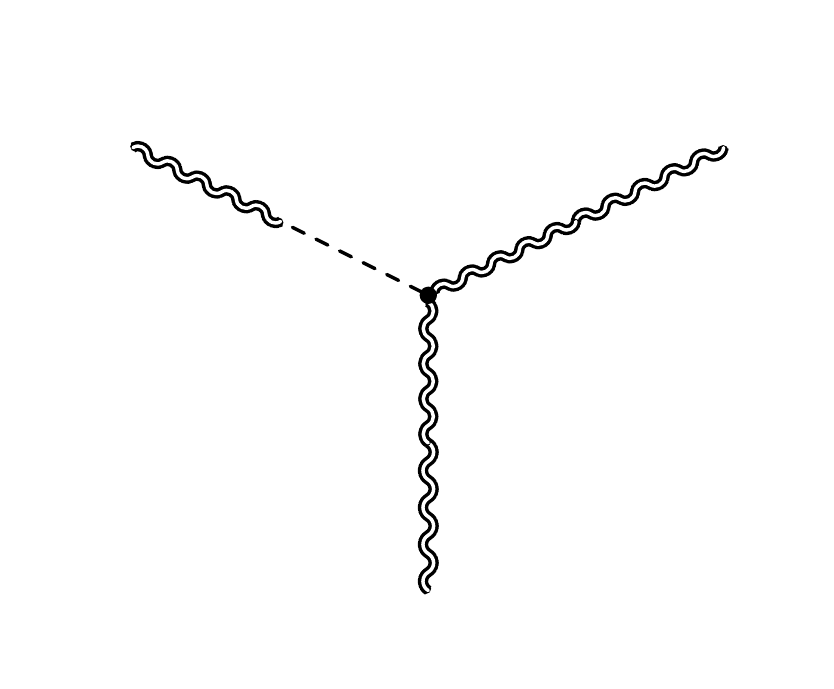} 
\includegraphics[scale=0.45]{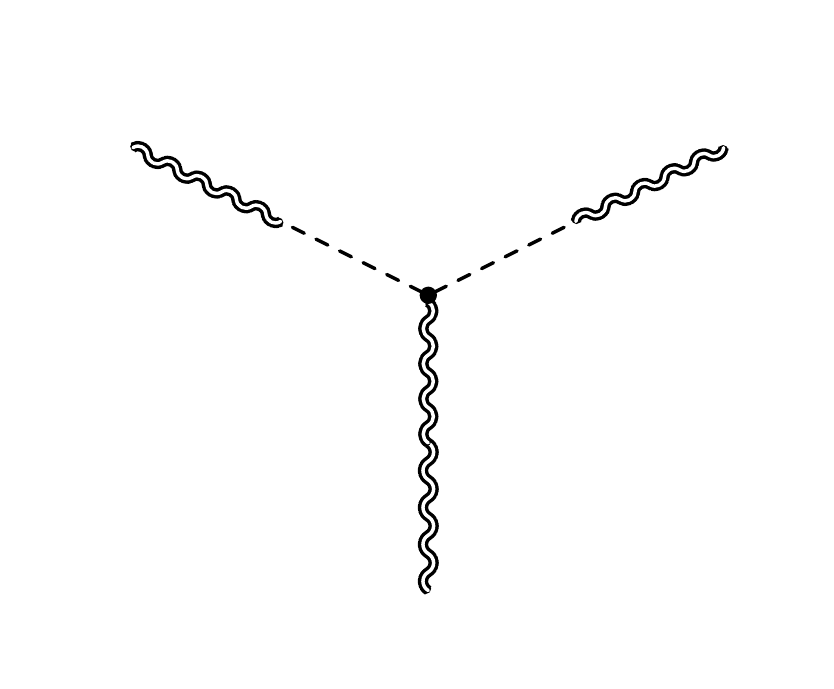} 
\includegraphics[scale=0.45]{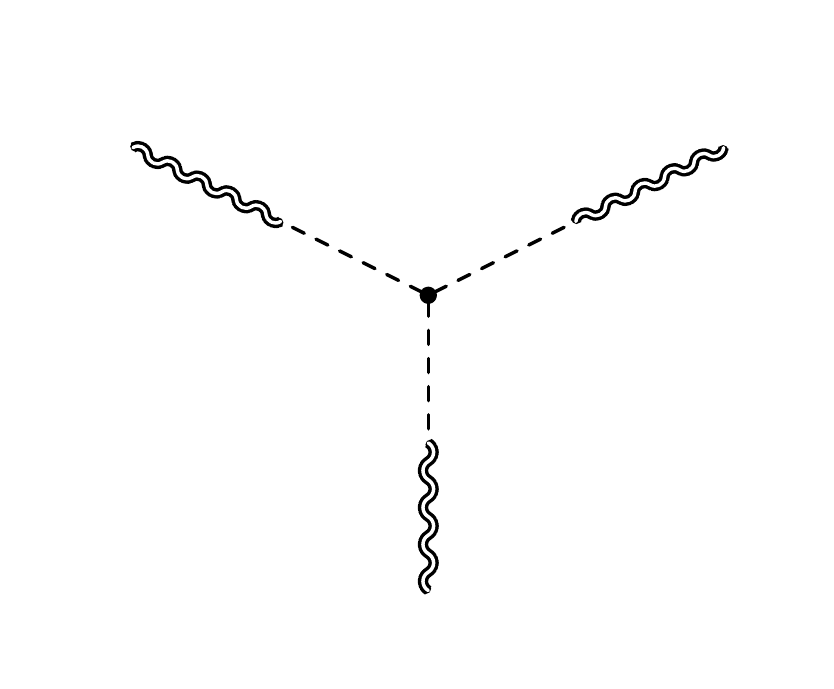} 

\caption{Mixing in three-point functions}
\end{figure}
Notice that the expression above is purely polynomial since $p_2^2 \pi^{\m_2\n_2}(p_2)$ is a local term.
Thus in the first line of (\ref{AS3}) we may substitute (\ref{traceS3}) and in the second line use (\ref{dbltrace3}) 
to eliminate the $Q^{\m_2\n_2}$ terms and its three cyclic permutations. The triple trace gives
\begin{align}
&\delta_{\a_1\b_1}\delta_{\a_2\b_2}\delta_{\a_3\b_3}\,S_3^{\a_1\b_1\a_2\b_2\a_3\b_3}(p_1,p_2,p_3)\big\vert_{p_3 = -(p_1 + p_2)} =\nonumber\\
&16\alpha \left[ p_1^2\,p_2^2 - (p_1\cdot p_2)^2\right].
\label{triptrace3}
\end{align}
We use these expressions to derive the structure of the 3-wave interaction in the form
\begin{align}
&S_3^{\m_1\n_1\m_2\n_2\m_3\n_3}=\sdfrac{1}{3}\, \pi^{\m_1\n_1}(p_1)\,\delta_{\a_1\b_1}\,S_3^{\a_1\b_1\m_2\n_2\m_3\n_3}+\nonumber\\
&\sdfrac{1}{3}\, \pi^{\m_2\n_2}(p_2)\,\delta_{\a_2\b_2}\,S_3^{\m_1\n_1\a_2\b_2\m_3\n_3}\nonumber\\
&+\sdfrac{1}{3}\, \pi^{\m_3\n_3}(p_3)\,\delta_{\a_3\b_3}\,S_3^{\m_1\n_1\m_2\n_2\a_3\b_3}- \nonumber\\
&\sdfrac{1}{9}\,\pi^{\m_1\n_1}(p_1)\,\pi^{\m_3\n_3}(p_3)\,\delta_{\a_1\b_1}\delta_{\a_3\b_3}\,S_3^{\a_1\b_1\m_2\n_2\a_3\b_3}- \nonumber\\
&\sdfrac{1}{9}\,\pi^{\m_2\n_2}(p_2)\pi^{\m_3\n_3}(p_3)\,\delta_{\a_2\b_2}\delta_{\a_3\b_3}\,S_3^{\m_1\n_1\a_2\b_2\a_3\b_3}-\nonumber\\
&\sdfrac{1}{9}\,\pi^{\m_1\n_1}(p_1)\pi^{\m_2\n_2}(p_2)\delta_{\a_1\b_1}\, \delta_{\a_2\b_2}\,S_3^{\a_1\b_1\a_2\b_2\m_3\n_3}+\nonumber\\
&\sdfrac{1}{27}\,\pi^{\m_1\n_1}(p_1)\pi_2^{\m_2\n_2} (p_2)\pi^{\m_3\n_3}(p_3)\,\delta_{\a_1\b_1}\delta_{\a_2\b_2}\delta_{\a_3\b_3}\,\nonumber\\
&S_3^{\a_1\b_1\a_2\b_2\a_3\b_3}\,.
\label{fin1}
\end{align}
The nonlocal $EGB$ theory has a structure that at trilinear level in the fluctuations, similarly to the case of the nonlocal 
anomaly actions, can be depicted as in Fig. 1. The vertex of the 3-wave is organized in terms of longitudinal insertions of massless states on each of the weavy lines, in a sequence of single, double and triple insertions. The dark blob at the center denotes the polynomial contributions coming from the functional derivatives of the Euler-Poincar\` e density. Each of the $\pi^{\mu\nu}$ projectors introduces a massless pole in momentum space, identified from the expression 
\beq
\pi^{\mu\nu}=\frac{1}{p^2}\hat{\pi}^{\mu\nu}\qquad \hat{\pi}^{\mu\nu}=\left(\delta^{\mu\nu}p^2 -p^\mu p^\nu\right)
\eeq
which induce nonlocal corrections on each of the external gravitational metric fluctuations $h_{\mu\nu}$. This picture gets modifed when we move to the case of 4-wave interactions. In that case the nonlocal action and henceforth Fig. 2 does not provide the correct expression of the vertices, and one has to resort to a perturbative expansion. While the specific features of these vertices  can always be obtained by the brute force expansion of the $V_E$ term around flat space to fourth order, there are some uncommon features that are typical of the expansion, as in the cubic case. The vertex can be reshuffled in an interesting way around flat space, as we are now going to show. 
\section{4-wave interactions in the subtracted vertex} 
Also in this case, as for conformal anomaly actions, 
\begin{figure}
\label{ff2}
\center
\includegraphics[scale=0.45]{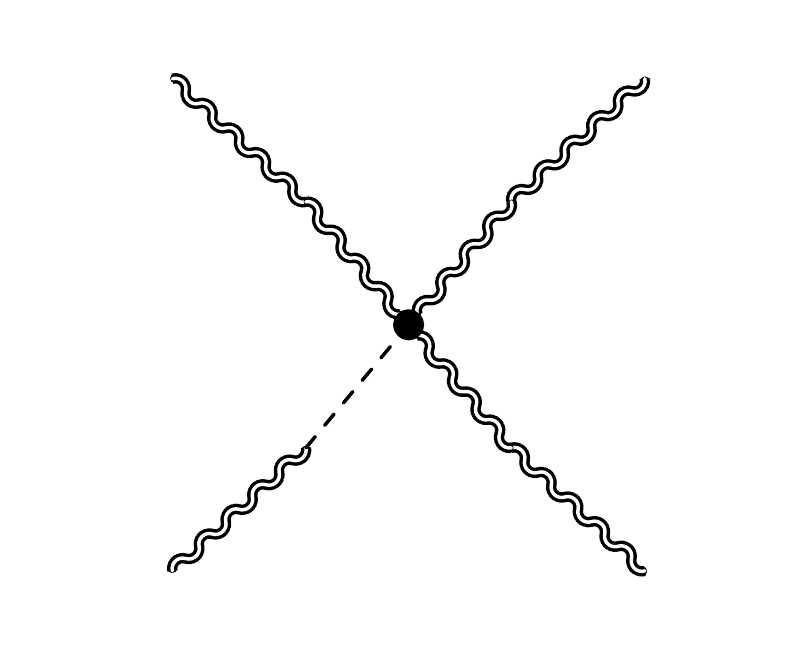} 
\includegraphics[scale=0.45]{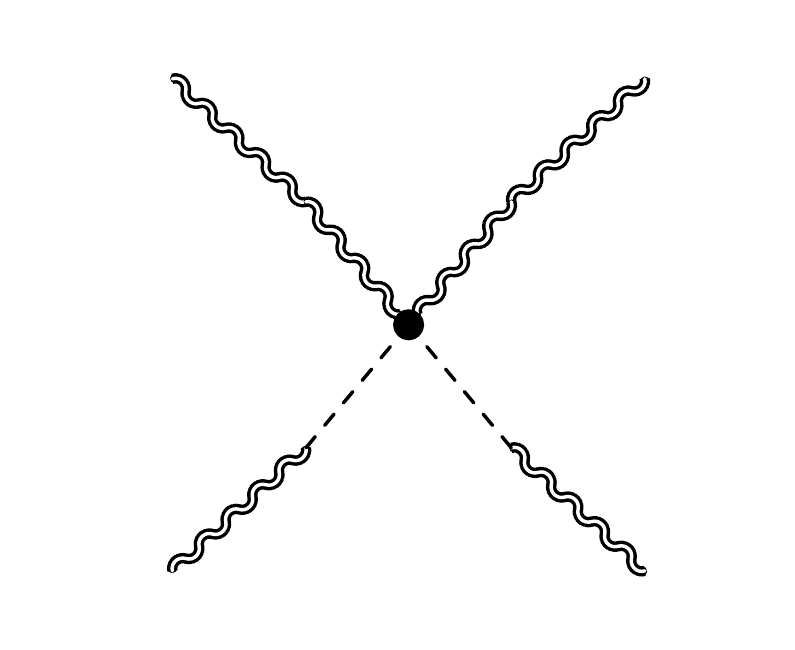} 
\includegraphics[scale=0.45]{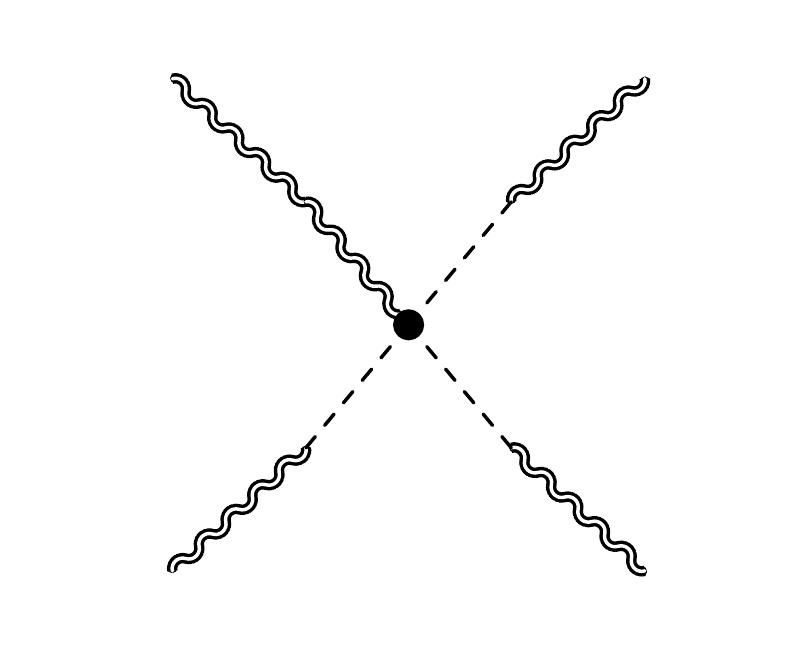} 
\includegraphics[scale=0.45]{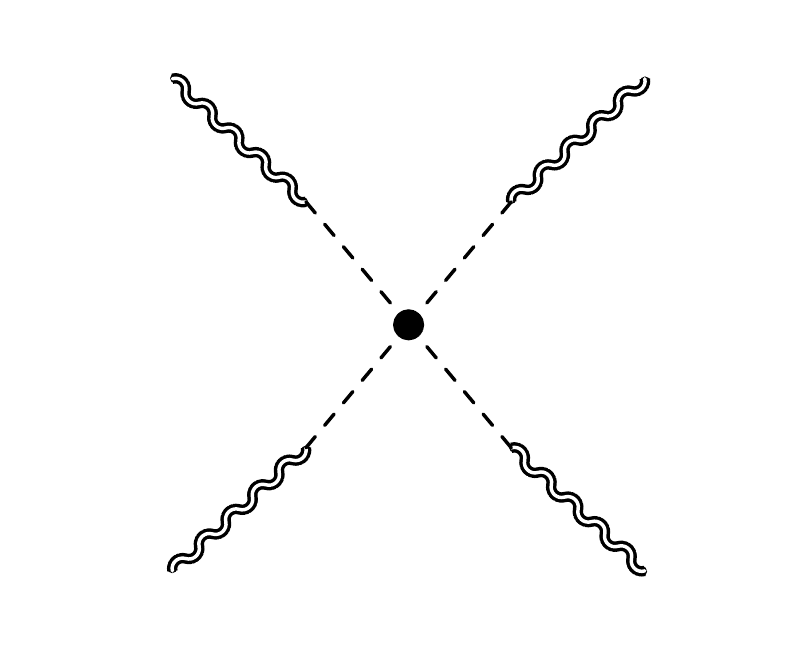} 
\caption{Mixing in four-point functions}
\end{figure}
we perform a subtraction of the form 
\begin{align}
&\mathcal{S}=\lim_{d\to4}\frac{1}{(d-4)}\left[V_E(g,d)-V_E(g,4)\right]=\nonumber\\
&\lim_{d\to4}\frac{1}{(d-4)}\left[\int\,d^dx\,\mu^{d-4}\,E_4-\int\,d^4x\,E_4\right]\label{diffE4}.
\end{align}
Notice that the subtraction introduced above differs by Weyl invariant terms from \eqref{WZ} 
\cite{Coriano:2022ftl,Coriano:2023sab} and it reproduces, in the flat spacetime limit, the usual regularization approach of DR with $\bar{g}\to \delta$, as applied, ordinarily, to momentum space. The subtraction is needed in order to account for the $0/0$ evanescent contributions that, as we have already mentioned, are present in the differentiation of $V_E$ with open indices.  The derivation of the conformal constraints, in this case, can be adapted from \cite{Coriano:2021nvn}, where it has been shown that these constraints can be directly 
extracted from the counterterms introduced to regulate the quantum corrections coming from a conformal sector. 
One may organize the contribution to the 4-point vertex (4-wave interaction) in terms of a sequence of anomaly poles and of a traceless contribution ("0-trace") derived from a consistent longitudinal/transverse decomposition of the 4-graviton interaction 
\begin{align}
&\mathcal{S}^{{\mu_1\nu_1}{\mu_2\nu_2}{\mu_3\nu_3}{\mu_4\nu_4}}(p_1,p_2,p_3,\bar{p}_4)=\nonumber\\
&\mathcal{S}^{{\mu_1\nu_1}{\mu_2\nu_2}{\mu_3\nu_3}{\mu_4\nu_4}}_{pole}(p_1,p_2,p_3,\bar{p}_4) +\nonumber\\
&\mathcal{S}^{{\mu_1\nu_1}{\mu_2\nu_2}{\mu_3\nu_3}{\mu_4\nu_4}}_{0-trace}(p_1,p_2,p_3,\bar{p}_4)\label{4S}.
\end{align}
the expression of $\mathcal{S}_{0-trace}$ is given below in \eqref{otrace}.
It is worth mentioning that the subtraction in \eqref{diffE4} is necessary in order to have a finite functional variation with respect to the metric fluctuation in the limit $d\to4$, as also pointed out in \cite{Bzowski:2017poo, Bzowski:2018fql}. This fact is reflected in the presence of an evanescent $0/0$ piece, due to Lovelock tensor identities, that can be eliminated once the subtraction \eqref{diffE4} is taken into account. There is a difference in the handling expressions such as \eqref{diffE4}, due to this subtle behaviour, if we perform functional derivatives with respect to the metric either with uncontracted or contracted indices. For instance, a differentiation with respect to the $\phi$ commutes with the limit in \eqref{diffE4}, and indeed reproduces the anomaly contribution in \eqref{wwi}, but a differentiation with open indices, followed by the flat Minkowski limit  $g\to\delta$, needs special care. If we include the subtraction given in  \eqref{diffE4} we can follow in the $4d$ EGB the same approach implemented in the case of a conformal anomaly action. 
In \eqref{4S} the zero trace part has the property 
\begin{align}
	&\delta_{\mu_i\nu_i}\mathcal{S}^{{\mu_1\nu_1}{\mu_2\nu_2}{\mu_3\nu_3}{\mu_4\nu_4}}_{0-trace}(p_1,p_2,p_3,\bar{p}_4)=0,\nonumber\\
	& i=1,2,3,4,
\end{align}
The explicit expression of $\mathcal{S}_{0-trace}$ requires the inclusion of suitable projectors separating the tensorial indices into transverse traceless, trace and longitudinal sectors, introduced in the analysis of the CWIs for 4T correlators in momentum space.  This contribution, in the context of the action functional \eqref{diffE4}, and  more generally, of conformally anomaly actions, has been shown to be necessary in order to generate consistent conservation WIs, as also discussed in \cite{Coriano:2022jkn}. We have  
\begin{align}
	&\mathcal{S}^{{\mu_1\nu_1}{\mu_2\nu_2}{\mu_3\nu_3}{\mu_4\nu_4}}(p_1,p_2,p_3,\bar{p}_4)_{0-trace}=\notag\\
	&=\bigg\{\mathcal{I}^{\mu_1\nu_1}_{\alpha_1}(p_1)\mathcal{I}^{\mu_2\nu_2}_{\alpha_2}(p_2)\mathcal{I}^{\mu_3\nu_3}_{\alpha_3}(p_3)\nonumber\\
	&\mathcal{I}^{\mu_4\nu_4}_{\alpha_4}(p_4)\,p_{1\beta_1}\,p_{2\beta_2}\,p_{3\beta_3}\,p_{4\beta_4}\,\notag\\
	&\qquad+\bigg[\Pi^{\mu_1\nu_1}_{\alpha_1\beta_1}(p_1)\mathcal{I}^{\mu_2\nu_2}_{\alpha_2}(p_2)\mathcal{I}^{\mu_3\nu_3}_{\alpha_3}(p_3)\nonumber\\
	&\mathcal{I}^{\mu_4\nu_4}_{\alpha_4}(p_4)\,p_{2\beta_2}\,p_{3\beta_3}\,p_{4\beta_4}+(\text{perm.})\bigg]\notag\\
	&\qquad+\bigg[\Pi^{\mu_1\nu_1}_{\alpha_1\beta_1}(p_1)\Pi^{\mu_2\nu_2}_{\alpha_2\beta_2}(p_2)\mathcal{I}^{\mu_3\nu_3}_{\alpha_3}(p_3)\nonumber\\
	&\mathcal{I}^{\mu_4\nu_4}_{\alpha_4}(p_4)\,p_{3\beta_3}\,p_{4\beta_4}+(\text{perm.})\bigg]\notag\\
	&\qquad+\bigg[\Pi^{\mu_1\nu_1}_{\alpha_1\beta_1}(p_1)\Pi^{\mu_2\nu_2}_{\alpha_2\beta_2}(p_2)\Pi^{\mu_3\nu_3}_{\alpha_3\beta_3}(p_3)\nonumber\\
	&\mathcal{I}^{\mu_4\nu_4}_{\alpha_4}(p_4)\,p_{4\beta_4}+(\text{perm.})\bigg]\bigg\}\nonumber\\
	&\mathcal{S}_{4}^{\alpha_1\beta_1\alpha_2\beta_2\alpha_3\beta_3\alpha_4\beta_4}(p_1,p_2,p_3,\bar{p}_4),
\label{otrace}
\end{align}
where the contractions are expressed in terms of the contribution of the third and second functional derivatives as
	\begin{align}
		&p_{1\,\beta_1}\,\mathcal{S}_{4}^{\alpha_1\beta_1\alpha_2\beta_2\alpha_3\beta_3\alpha_4\beta_4}(p_1,p_2,p_3,\bar{p}_4)=\notag\\
		&=\Big[4\, \mathcal{B}^{\alpha_1\hspace{0.4cm}\alpha_2\beta_2\alpha_3\beta_3}_{\hspace{0.3cm}\lambda_1\beta_1}(p_2,p_3)\,\nonumber\\
		&\mathcal{S}_{2}^{\lambda_1\beta_1\alpha_4\beta_4}(p_1+p_2+p_3,\bar{p}_4)+(34)+ (24)\Big]\notag\\
		&\hspace{0.9cm}+\Big[2 \, \mathcal{C}^{\alpha_1\hspace{0.4cm}\alpha_2\beta_2}_{\hspace{0.3cm}\lambda_1\beta_1}(p_2)\,\nonumber\\
		&\mathcal{S}_{3}^{\lambda_1\beta_1\alpha_3\beta_3\alpha_4\beta_4}(p_1+p_2,p_3,\bar{p}_4)+(2 3)+(24)\Big].
	\end{align} 
We have defined the transverse traceless projector 
\begin{align}
\Pi^{\mu \nu}_{\alpha \beta} & = \frac{1}{2} \left( \pi^{\mu}_{\alpha} \pi^{\nu}_{\beta} + \pi^{\mu}_{\beta} \pi^{\nu}_{\alpha} \right) - \frac{1}{d - 1} \pi^{\mu \nu}\pi_{\alpha \beta}\label{TTproj}, 
\end{align}
and the tensor
\begin{align}
\mathcal{I}^{\mu\nu}_{\alpha}&=\frac{1}{p^2}\left( p^{\mu}\delta^{\nu}_{\alpha} +
p^{\nu}\delta^{\mu}_{\alpha} -
\frac{p_{\alpha}}{d-1}( \delta^{\mu\nu} +(d-2)\frac{p^\mu p^\nu}{p^2}  \right)\label{proone}.
\end{align}
On the other hand, the pole part is then explicitly given as
\begin{align}
	&\mathcal{S}^{{\mu_1\nu_1}{\mu_2\nu_2}{\mu_3\nu_3}{\mu_4\nu_4}}(p_1,p_2,p_3,\bar{p}_4)_{poles}=\notag\\
	&=\frac{\pi^{\mu_1\nu_1}(p_1)}{3}\delta_{\alpha_1\beta_1}\mathcal{S}_{4}^{\alpha_1\beta_1\mu_2\nu_2\mu_3\nu_3\mu_4\nu_4}(p_1,p_2,p_3,\bar{p}_4)+\nonumber\\
	&(perm.)-\frac{\pi^{\mu_1\nu_1}(p_1)}{3}\,\frac{\pi^{\mu_2\nu_2}(p_2)}{3}\,\delta_{\alpha_1\beta_1}\delta_{\alpha_2\beta_2}\times\nonumber\\
	&\times\mathcal{S}_{4}^{\alpha_1\beta_1\alpha_2\beta_2\mu_3\nu_3\mu_4\nu_4}(p_1,p_2,p_3,\bar{p}_4)+(perm.)\notag\\
	&+\frac{\pi^{\mu_1\nu_1}(p_1)}{3}\,\frac{\pi^{\mu_2\nu_2}(p_2)}{3}\,\frac{\pi^{\mu_3\nu_3}(p_3)}{3}\,\delta_{\alpha_2\beta_2}\,\delta_{\alpha_1\beta_1}\delta_{\alpha_2\beta_2}\delta_{\alpha_3\beta_3}\,\nonumber\\
	&\mathcal{S}_{4}^{\alpha_1\beta_1\alpha_2\beta_2\alpha_3\beta_3\mu_4\nu_4}(p_1,p_2,p_3,\bar{p}_4)+(perm.)\notag\\
	&-\frac{\pi^{\mu_1\nu_1}(p_1)}{3}\,\frac{\pi^{\mu_2\nu_2}(p_2)}{3}\,\frac{\pi^{\mu_3\nu_3}(p_3)}{3}\,\frac{\pi^{\mu_4\nu_4}(p_4)}{3}\times\,\nonumber\\
&\times\delta_{\alpha_1\beta_1}\delta_{\alpha_2\beta_2}\delta_{\alpha_3\beta_3}\delta_{\alpha_4\beta_4}\mathcal{S}_{4}^{\alpha_1\beta_1\alpha_2\beta_2\alpha_3\beta_3\alpha_4\beta_4}(p_1,p_2,p_3,\bar{p}_4),
\end{align}
where the single trace is expressed in terms of the third functional derivative  
 \begin{align}
	&\delta_{\alpha_1\beta_1}\,\mathcal{S}_4^{\alpha_1\beta_1\alpha_2\beta_2\alpha_3\beta_3\alpha_4\beta_4}(p_1,p_2,p_3,p_4)=\nonumber\\
	&8\,\left[\sqrt{-g}E\right]^{\alpha_2\beta_2\alpha_3\beta_3\alpha_4\beta_4}(p_2,p_3,p_4)\notag\\
	&-2\bigg[\mathcal{S}_{3}^{\alpha_2\beta_2\alpha_3\beta_3\alpha_4\beta_4}(p_1+p_2,p_3,p_4)+\nonumber\\
	&\mathcal{S}_{3}^{\alpha_2\beta_2\alpha_3\beta_3\alpha_4\beta_4}(p_2,p_1+p_3,p_4)+\nonumber\\
	&\mathcal{S}_{3}^{\alpha_2\beta_2\alpha_3\beta_3\alpha_4\beta_4}(p_2,p_3,p_1+p_4)\bigg],
\end{align}
and additional traces involve vertices of lower orders. 

\section{Conclusions}
In this work we have investigated the nonlinear constraints emerging in a nonlocal 4d EGB theory for the 
3- and 4-point classical interactions present in its fundamental action. They are hierarchical and directly linked to 
to the topological properties of the vertex $V _E$ and its subtracted expression $\hat{V}'_E$, valid in $d\neq 4$ and $d=4$ dimensions respectively. \\
The analysis is essentially built on several previous studies of conformal anomaly actions, that allow to identify some nontrivial features of this specific $4d$ theory. Among these, the presence of bilinear mixing 
in the graviton vertices and of extra, traceless contributions, which are not predicted by the nonlocal action as we move to 4-wave interactions. 
The nonlocal version of such a theory, as pointed out, is derived by a finite renormalization of the topological density, that allows to remove the dilaton from the spectrum by expressing it in terms of the full metric. Three-wave interactions are naturally derived from the nonlocal action, but the derivation of the hierarchical constraint satisfied by the four-wave interactions requires a different approach, 
given the limitations of such actions in reproducing the correct flat spacetime limit. The constraints that we have derived are satisfied also in the case of Lovelock actions. Indeed it is possible to generalize them to those cases where topological invariants of higher orders, such as $E_6, E_8$ and so on, are present, 
extending the approach outlined in this work. We hope to return to this point in a future study.  

\centerline{\bf Acknowledgements}
We thank Riccardo Tommasi for discussions.
This work is partially supported by INFN within the Iniziativa Specifica QFT-HEP.  
The work of C. C. and M. C. is funded by the European Union, Next Generation EU, PNRR project "National Centre for HPC, Big Data and Quantum Computing", project code CN00000013 and by INFN iniziativa specifica QFT-HEP. The work of M.C. is also supported by a PON fellowship DOT1312457-3.
M. M. M. is supported by the European Research Council (ERC) under the European Union as Horizon 2020 research and innovation program (grant agreement No818066) and by Deutsche Forschungsgemeinschaft (DFG, German Research Foundation) under Germany's Excellence Strategy EXC-2181/1 - 390900948 (the Heidelberg STRUCTURES Cluster of Excellence).    

%\bibliographystyle{jhep}
%\bibliography{TJJdilatonH3}

\begin{thebibliography}{10}

\bibitem{Duff:1986pq}
M.~J. Duff, B.~E.~W. Nilsson, and C.~N. Pope, {\it {{Gauss-Bonnet} From
  {Kaluza-Klein}}},  {\em Phys. Lett. B} {\bf 173} (1986) 69--72.

\bibitem{Zwiebach:1985uq}
B.~Zwiebach, {\it {Curvature Squared Terms and String Theories}},  {\em Phys.
  Lett. B} {\bf 156} (1985) 315--317.

\bibitem{Glavan:2019inb}
D.~Glavan and C.~Lin, {\it {Einstein-Gauss-Bonnet Gravity in Four-Dimensional
  Spacetime}},  {\em Phys. Rev. Lett.} {\bf 124} (2020), no.~8 081301,
  [\href{http://xxx.lanl.gov/abs/1905.0360}{{\tt arXiv:1905.0360}}].

\bibitem{Mann:1992ar}
R.~B. Mann and S.~F. Ross, {\it {The D ---\ensuremath{>} 2 limit of general
  relativity}},  {\em Class. Quant. Grav.} {\bf 10} (1993) 1405--1408,
  [\href{http://xxx.lanl.gov/abs/gr-qc/9208004}{{\tt gr-qc/9208004}}].

\bibitem{Coriano:2023sab}
C.~Corian\`o, M.~Cret\`\i{}, and M.~M. Maglio, {\it {Broken Scale Invariance
  and the Regularization of a Conformal Sector in Gravity with Wess-Zumino
  actions}},  \href{http://xxx.lanl.gov/abs/2301.0746}{{\tt arXiv:2301.0746}}.

\bibitem{Matsumoto:2022fln}
M.~Matsumoto and Y.~Nakayama, {\it {Dilaton invading from infinitesimal extra
  dimension}},  \href{http://xxx.lanl.gov/abs/2202.1353}{{\tt
  arXiv:2202.1353}}.

\bibitem{Coriano:2013nja}
C.~Corian\`o, L.~Delle~Rose, C.~Marzo, and M.~Serino, {\it {The dilaton
  Wess-Zumino action in six dimensions from Weyl gauging: local anomalies and
  trace relations}},  {\em Class. Quant. Grav.} {\bf 31} (2014) 105009,
  [\href{http://xxx.lanl.gov/abs/1311.1804}{{\tt arXiv:1311.1804}}].

\bibitem{Ferreira:2017wqz}
F.~M. Ferreira and I.~L. Shapiro, {\it {Integration of trace anomaly in 6D}},
  {\em Phys. Lett. B} {\bf 772} (2017) 174--178,
  [\href{http://xxx.lanl.gov/abs/1702.0689}{{\tt arXiv:1702.0689}}].

\bibitem{Elvang:2012st}
H.~Elvang, D.~Z. Freedman, L.-Y. Hung, M.~Kiermaier, R.~C. Myers, {\em
  et.~al.}, {\it {On renormalization group flows and the a-theorem in 6d}},
  {\em JHEP} {\bf 1210} (2012) 011,
  [\href{http://xxx.lanl.gov/abs/1205.3994}{{\tt arXiv:1205.3994}}].


\bibitem{Lovelock:1971yv}
D.~Lovelock, {\it {The Einstein tensor and its generalizations}},  {\em J.
  Math. Phys.} {\bf 12} (1971) 498--501.

\bibitem{Coriano:2022ftl}
C.~Corian\`o, M.~M. Maglio, and D.~Theofilopoulos, {\it {Topological
  corrections and conformal backreaction in the Einstein
  Gauss\textendash{}Bonnet/Weyl theories of gravity at $D=4$}},  {\em Eur.
  Phys. J. C} {\bf 82} (2022), no.~12 1121,
  [\href{http://xxx.lanl.gov/abs/2203.0421}{{\tt arXiv:2203.0421}}].

\bibitem{Edelstein:2014dje}
J.~D. Edelstein, {\it {Lovelock theory, black holes and holography}},  {\em
  Springer Proc. Math. Stat.} {\bf 60} (2014) 19--36,
  [\href{http://xxx.lanl.gov/abs/1303.6213}{{\tt arXiv:1303.6213}}].

\bibitem{Charmousis:2014mia}
C.~Charmousis, {\it {From Lovelock to Horndeski`s Generalized Scalar Tensor
  Theory}},  {\em Lect. Notes Phys.} {\bf 892} (2015) 25--56,
  [\href{http://xxx.lanl.gov/abs/1405.1612}{{\tt arXiv:1405.1612}}].

\bibitem{Mazur:2001aa}
P.~O. Mazur and E.~Mottola, {\it {Weyl cohomology and the effective action for
  conformal anomalies}},  {\em Phys.Rev.} {\bf D64} (2001) 104022,
  [\href{http://xxx.lanl.gov/abs/hep-th/0106151}{{\tt hep-th/0106151}}].

\bibitem{Barvinsky:1995it}
A.~O. Barvinsky, A.~G. Mirzabekian, and V.~V. Zhytnikov, {\it {Conformal
  decomposition of the effective action and covariant curvature expansion}},
  in {\em {6th Moscow Quantum Gravity}}, 6, 1995.
\newblock \href{http://xxx.lanl.gov/abs/gr-qc/9510037}{{\tt gr-qc/9510037}}.

\bibitem{Coriano:2022jkn}
C.~Corian\`o, M.~M. Maglio, and R.~Tommasi, {\it {Four-point functions of
  gravitons and conserved currents of CFT in momentum space: testing the
  nonlocal action with the TTJJ}},
  \href{http://xxx.lanl.gov/abs/2212.1277}{{\tt arXiv:2212.1277}}.

\bibitem{Coriano:2021nvn}
C.~Corian\`o, M.~M. Maglio, and D.~Theofilopoulos, {\it {The Conformal Anomaly
  Action to Fourth Order (4T) in $d=4$ in Momentum Space}},
  \href{http://xxx.lanl.gov/abs/2103.1395}{{\tt arXiv:2103.1395}}.

\bibitem{Coriano:2022knl}
C.~Corian\`o and M.~M. Maglio, {\it {Einstein Gauss-Bonnet theories as
  ordinary, Wess-Zumino conformal anomaly actions}},  {\em Phys. Lett. B} {\bf
  828} (2022) 137020, [\href{http://xxx.lanl.gov/abs/2201.0751}{{\tt
  arXiv:2201.0751}}].

\bibitem{Coriano:2017mux}
C.~Corian\`o, M.~M. Maglio, and E.~Mottola, {\it {TTT in CFT: Trace Identities
  and the Conformal Anomaly Effective Action}},  {\em Nucl. Phys.} {\bf B942}
  (2019) 303--328, [\href{http://xxx.lanl.gov/abs/1703.0886}{{\tt
  arXiv:1703.0886}}].

\bibitem{Bzowski:2017poo}
A.~Bzowski, P.~McFadden, and K.~Skenderis, {\it {Renormalised 3-point functions
  of stress tensors and conserved currents in CFT}},
  \href{http://xxx.lanl.gov/abs/1711.0910}{{\tt arXiv:1711.0910}}.

\bibitem{Bzowski:2018fql}
A.~Bzowski, P.~McFadden, and K.~Skenderis, {\it {Renormalised CFT 3-point
  functions of scalars, currents and stress tensors}},  {\em JHEP} {\bf 11}
  (2018) 159, [\href{http://xxx.lanl.gov/abs/1805.1210}{{\tt
  arXiv:1805.1210}}].

\end{thebibliography}
 
\providecommand{\href}[2]{#2}\begingroup\raggedright\endgroup

 \end{document}